\documentclass[aps,prxquantum,twocolumn,english,superscriptaddress,floatfix,longbibliography]{revtex4-2}
\usepackage[english]{babel}
\usepackage{textcomp,xcolor,natbib}
\usepackage{dcolumn}
\usepackage{graphicx}
\graphicspath{{Figures/}}

\usepackage{amsmath, bbm}
\usepackage{latexsym}
\usepackage{amsfonts}   
\usepackage{amssymb}
\usepackage{array}      
\usepackage{epsfig}
\usepackage{txfonts}
\usepackage{xcolor,braket}
\usepackage[colorlinks=true,linkcolor=blue,urlcolor=blue,citecolor=blue]{hyperref}
\usepackage[normalem]{ulem}
\usepackage{soul}
\usepackage{dsfont}

\usepackage{xfrac}  
\usepackage{relsize}

\usepackage{amsthm}
\usepackage{mathrsfs, mathtools, amsmath}

\usepackage[normalem]{ulem}
\usepackage{cancel}
\usepackage{enumitem}

\usepackage{xspace}
\usepackage{enumitem}

\newcommand{\ketbra}[1]{{\ket{#1}\bra{#1}}}
\newcommand{\id}{{\mathbbm{1}}}
\newcommand{\abs}[1]{\left\lvert #1 \right\rvert}
\newcommand{\norm}[1]{\left\lVert #1 \right\rVert}
\newcommand{\tr}{{\operatorname{tr}}}

\newcommand{\schro}{Schr\"odinger\xspace}
\newcommand{\heis}{Heisenberg\xspace}

\newtheorem{proposition}{Proposition}

\newtheorem{corollary}{Corollary}

\newtheorem{example}{Example}

\begin{document}
\title{Divisibility of dynamical maps: Schr\"odinger vs. Heisenberg picture}

\author{Federico Settimo}
\email{fesett@utu.fi}
\affiliation{Department of Physics and Astronomy,
University of Turku, FI-20014 Turun yliopisto, Finland}
%\orcid{0000-0002-0123-6950}

\author{Andrea Smirne}
\affiliation{Dipartimento di Fisica ``Aldo Pontremoli'', Universit{\`a} degli Studi di Milano, Via Celoria 16, I-20133 Milan, Italy}
\affiliation{Istituto Nazionale di Fisica Nucleare, Sezione di Milano, Via Celoria 16, I-20133 Milan, Italy}
%\orcid{0000-0003-4698-9304}

\author{Kimmo Luoma}
\affiliation{Department of Physics and Astronomy,
University of Turku, FI-20014 Turun yliopisto, Finland}
%\orcid{0000-0003-3118-612X}

\author{Bassano Vacchini}
\affiliation{Dipartimento di Fisica ``Aldo Pontremoli'', Universit{\`a} degli Studi di Milano, Via Celoria 16, I-20133 Milan, Italy}
\affiliation{Istituto Nazionale di Fisica Nucleare, Sezione di Milano, Via Celoria 16, I-20133 Milan, Italy}
%\orcid{0000-0002-7574-9951}

\author{Jyrki Piilo}
\affiliation{Department of Physics and Astronomy,
University of Turku, FI-20014 Turun yliopisto, Finland}
%\orcid{0000-0002-5595-873X}

\author{Dariusz Chru\'sci\'nski}
\affiliation{Institute of Physics, Faculty of Physics, Astronomy and Informatics,
Nicolaus Copernicus University, Grudziadzka 5/7, 87-100 Toru\'{n},
Poland}
%\orcid{0000-0002-6582-6730}

\begin{abstract}
    Divisibility of dynamical maps is a central notion in the study of quantum non-Markovianity, providing a natural framework to characterize memory effects via time-local master equations.
    In this work, we generalize the notion of divisibility of quantum dynamical maps from the Schr\"odinger to the Heisenberg picture.
    While the two pictures are equivalent at the level of physical predictions, we show that the divisibility properties of the corresponding dual maps are, in general, not equivalent.
    This inequivalence originates from the distinction between left and right generators of time-local master equations, which interchange roles under duality.
    We demonstrate that Schr\"odinger and Heisenberg divisibility are distinct concepts by constructing explicit dynamics divisible only in one picture.
    Furthermore, we introduce a quantifier for the violation of Heisenberg P-divisibility, analogous to the trace-distance-based measure of non-Markovianity, and provide it with an operational interpretation in terms of the guessing probability between effects.
    Our results show that Heisenberg divisibility is an independent witness of memory effects and highlight the need to consider both pictures when characterizing non-Markovian quantum dynamics.
\end{abstract}

\maketitle

% ------------------------------------- Intro -------------------------------------
\section{Introduction}
\label{sec:intro}

The concept of divisibility for open quantum system dynamics has been widely investigated, in particular in connection with non-Markovianity.
A quantum evolution, represented by a dynamical map $\{\Phi_t\}_{t\geq 0}$, is said to be P- or CP-divisible if the evolution between any two times $s$ and $t > s$  can be described by positive (PTP) or completely positive trace preserving (CPTP) maps $\Phi_{t,s}$. In other words, the dynamical map at time $t$ can be {written} as a composition of (C)PTP maps $\Phi_{t_i,t_{i-1}}$
\begin{equation}
    \Phi_t = \Phi_{t,t_{n}} \circ \Phi_{t_{n-1},t_{n-2}} \circ \ldots \circ \Phi_{t_2,t_{1}} \circ \Phi_{t_1} ,
\end{equation}
for an arbitrary sequence $t \geq t_n \geq t_{n-1} \geq \ldots \geq t_1$.

Violations of divisibility have been connected to non-Markovianity and information backflow \cite{rivas-quantum-nm, BLPV-colloquium, Chruscinski2011, Wißmann-BLP-P-div, Buscemi2016, Li2019}.
If divisibility holds, then information can only flow from the system into the environment, while its violation can be interpreted as memory effects.
Other definitions of non-Markovianity based on the non-monotonicity of distinguishability quantifiers between states \cite{BLP, BLP-PRA, Chruscinski2011, Wißmann-BLP-P-div, Megier2021, Smirne2022, Settimo-JSD} or of entanglement \cite{RHP} have been proposed and some of them have been proved to be equivalent to violations of P- \cite{Chruscinski2011, Wißmann-BLP-P-div} or CP-divisibility \cite{RHP}.

Traditionally, the concept of divisibility has been only investigated in the \schro picture.
When the dynamics is governed by a time-local master equation (ME), divisibility can be linked to the properties of the time-dependent generator of the ME \cite{Kossakowski-necessary}, and its analysis provides insights into phenomena such as information backflow and memory effects.
Due to the fundamental duality between the \schro and \heis pictures, it is natural to ask how these concepts translate when moving from the evolution of states to the evolution of observables.
 
When considering semigroup dynamics \cite{Alicki2007}, the corresponding generator $\mathcal{L}$ is time-independent and hence $\Phi_t = e^{t \mathcal{L}}$. Note that $\dot{\Phi}_t = \mathcal{L} \circ \Phi_t = \Phi_t \circ \mathcal{L}$ due to the fact that $\mathcal{L}$ and $\Phi_t$ commute. Hence, it is not essential whether in the ME the generator is on the left  $\mathcal{L} \circ \Phi_t$ or on the right $\Phi_t \circ \mathcal{L}$.
Equivalently in the \heis picture $\mathcal{L}^*$ is simply the Hilbert-Schmidt adjoint of the generator $\mathcal{L}$ in the \schro picture. Again, one has
$\dot{\Phi}^*_t = \mathcal{L}^* \circ \Phi^*_t = \Phi^*_t \circ \mathcal{L}^*$. 
However, this is no longer true beyond a semigroup scenario when the generator is time-dependent and commutativity is no longer guaranteed.
Traditionally, one considers MEs $\dot{\Phi}_t = \mathcal{L}_t \circ \Phi_t$, but the map $\Phi_t$ satisfies also  $\dot{\Phi}_t =  \Phi_t \circ \mathcal{R}_t$ with an appropriate generator $\mathcal{R}_t$, that is generally different from $\mathcal{L}_t$, $\mathcal{R}_t \neq \mathcal{L}_t$. In this paper we propose to call a {\em traditional} $\mathcal{L}_t$ a {\em left generator} and  $\mathcal{R}_t$ a {\em right generator}. Of course, in the commutative case, i.e. when $\mathcal{L}_t$ commutes with $\Phi_t$, left and right generators coincide, $\mathcal{L}_t = \mathcal{R}_t$.

The meaning of the left generator $\mathcal{L}_t$ is clear. In particular it does control the divisibility property of the dynamical map $\Phi_t$ in the Schr\"odinger picture. What is then the meaning of the right generator $\mathcal{R}_t$?  
In this work, we generalize the concept of divisibility to the \heis picture and show that $\mathcal{R}_t^*$ controls the divisibility property of the dynamical map $\Phi^*_t$ in the Heisenberg  picture. 
Our key result is that traditional Schr\"odinger divisibility and Heisenberg divisibility are not equivalent, i.e. a dynamical map $\Phi_t$ may be Schr\"odinger divisible whereas $\Phi^*_t$ may be Heisenberg indivisible and vice versa.
Furthermore, we provide a quantifier of the violation of P-divisibility in the \heis picture analogous to the measure of non-Markovianity proposed in \cite{BLP, BLP-PRA}, connected to non-monotonicity of distinguishability quantifiers between effects.

The rest of the paper is organized as follows.
In Sec.~\ref{sec:left-right_gen}, we characterize the left and the right generators, both in the \schro and in the \heis picture, showing that a left generator becomes a right one when going from one picture to the other.
In Sec.~\ref{sec:divisibility}, we explore the concept of divisibility in both pictures, showing that in the \heis picture it is generated by the {\it right} generator of the \schro master equation.
We therefore conclude that divisibility in the two pictures are not equivalent and propose a measure of violations of divisibility in the \heis picture.
In Sec.~\ref{sec:compatibility}, we study \heis divisibility in connection to dynamical evolution of positive operator-valued measures (POVMs), showing that a non-monotonic behavior of either compatibility or sharpness can be used to witness violations of divisibility.
In Sec.~\ref{sec:qubit}, we propose an explicit example of qubit dynamics divisible only in one picture but not in the other.
Finally, in Sec.~\ref{sec:conclusions}, we provide an overview of our work and point to future developments.

% ------------------------------------- Left Right gen -------------------------------------
\section{Left and right generators of the master equation}
\label{sec:left-right_gen}

We now introduce the left and right generators for open quantum system dynamics, in both the \schro and the \heis picture.

\subsection{Schr\"odinger picture}

In open quantum systems, under the assumption that the open system is initially uncorrelated with its environment, the dynamics of the reduced density matrix is described by a CPTP dynamical map $\Phi_t:\rho\mapsto\Phi_t[\rho] = \rho(t)$ \cite{Breuer-Petruccione, Rivas-Huelga-OQS, Chruscinski2022, Vacchini2024}.
Assuming that the dynamics is invertible, i.e. that $\Phi_t^{-1}$ exists at all times, then $\Phi_t$ is guaranteed to be the solution of the time-local ME
\begin{equation}
    \label{eq:ME_left}
    \dot\Phi_t = \mathcal L_t\circ\Phi_t , %\qquad\Leftrightarrow\qquad \frac d{dt}\rho(t) = \mathcal % L_t\left[\Phi_t[\rho(0)]\right] = \mathcal L_t[\rho(t)],
\end{equation}
which, in turn, implies the following ME for the density operator
\begin{equation}
   \frac d{dt}\rho(t)  = \mathcal L_t[\rho(t)] .
\end{equation}
The time-dependent generator $\mathcal L_t = \dot\Phi_t\circ\Phi_t^{-1}$ has the following universal structure \cite{Gorini1976, Lindblad1976}
\begin{equation}
    \label{eq:Lind_left-rho}
    \begin{split}
        \mathcal L_t[\rho] = -&i[H(t),\rho]+\sum_\alpha \gamma_\alpha(t)\Big[L_\alpha(t)\rho L_\alpha^\dagger(t)\\
        &-\frac12\{L_\alpha^\dagger(t)L_\alpha(t),\rho\}\Big],
    \end{split}
\end{equation}
where $H(t) = H^\dagger(t)$, which follows from the requirements that the dynamics is trace- and Hermiticity-preserving, in turn implying
that the generator is trace-annihilating and Hermiticity-preserving, i.e. $\tr\mathcal L_t[\rho] = 0$ and $\mathcal L_t[\rho]^\dagger = \mathcal L_t[\rho]$.
The solution of the ME \eqref{eq:ME_left} can be written as
\begin{equation}
    \label{eq:Phi_time_ordered}
    \Phi_t = T_\leftarrow\exp\left(\int_0^td\tau\,\mathcal L_\tau\right),
\end{equation}
where $T_\leftarrow$ is the chronological time ordering.

Equivalently, the ME can be also written as
\begin{equation}
    \label{eq:ME_right}
    \dot\Phi_t = \Phi_t\circ\mathcal R_t ,
    %\qquad\Leftrightarrow\qquad \frac d{dt}\rho(t) = \Phi_t\left[\mathcal R_t[\rho(0)]\right],
\end{equation}
with an appropriate generator $\mathcal{R}_t$. Indeed, if one knows $\Phi_t$, then one can simply define $\mathcal R_t := \Phi_t^{-1}\circ\dot\Phi_t$. Note that $\mathcal{R}_t$ has the same form as $\mathcal{L}_t$ since it is trace-annihilating and Hermiticity-preserving as well, and therefore it can be represented as
\begin{gather}\notag
        \mathcal R_t [\rho] = -i[K(t),\rho] + \sum_\alpha\xi_\alpha(t)\Big[R_\alpha(t) \rho R_\alpha^\dagger(t)\\
        -\frac12\left\{R_\alpha^\dagger(t)R_\alpha(t),\rho\right\}\Big],
\end{gather}
with $K^\dagger(t) = K(t)$.
Therefore, the dynamical map $\Phi_t$ can alternatively be written as
\begin{equation}
    \label{eq:Phi_anti_time_ordered}
    \Phi_t = T_\rightarrow\exp\left(\int_0^td\tau\,\mathcal R_\tau\right),
\end{equation}
where $T_\rightarrow$ is the anti-chronological time ordering.
On the other hand, \eqref{eq:ME_right} does not imply a ME for $\rho(t)$,
but one has
\begin{equation}
     \frac d{dt}\rho(t) = \Phi_t\left[\mathcal R_t[\rho(0)]\right],
\end{equation}
which requires not only the knowledge of $\mathcal{R}_t$ but also of the map $\Phi_t$. 
However, knowing the map, one can immediately generate the corresponding state $\rho(t) = \Phi_t[\rho(0)]$ starting from the arbitrary initial state $\rho(0)$. This is the main reason why we traditionally use $\mathcal{L}_t$ but not $\mathcal{R}_t$: Knowing $\mathcal{L}_t$ and an initial state $\rho(0)$, we can find $\rho(t)$ by directly solving Eq.~\eqref{eq:Lind_left-rho}.

In the following, we refer to MEs of the form \eqref{eq:ME_left} as {\it left ME} and to its generator $\mathcal L_t$ as {\it left generator}.
Similarly, we refer to MEs of the form \eqref{eq:ME_right} as {\it right ME} and to generators $\mathcal R_t$ as {\it right generator}.
Naturally, it is possible to derive the left generator from the right one and viceversa (once the map $\Phi_t$ is known) as
\begin{equation}
    \label{eq:left_to_right_gen}
    \mathcal R_t = \Phi_t^{-1}\circ\mathcal L_t\circ\Phi_t,\qquad
    \mathcal L_t = \Phi_t\circ\mathcal R_t\circ \Phi^{-1}_t.
\end{equation}

By comparing Eqs.~\eqref{eq:Phi_time_ordered} and \eqref{eq:Phi_anti_time_ordered}, it is easy to see that in general $\mathcal R_t\ne\mathcal L_t$, unless
\begin{equation}
    \label{eq:commutativity}
    \left[\mathcal L_t,\mathcal L_{t^\prime}\right] = 0
\end{equation}
for all times $t$, $t^\prime$.
This, in turns, implies $[\mathcal L_t,\Phi_t]=0$ and using Eq.~\eqref{eq:left_to_right_gen} it implies $\mathcal L_t=\mathcal R_t$.
The commutativity of $\mathcal L_t$ occurs, for example, if the dynamics is a semigroup, for which $\mathcal L_t$ does not depend on time, or if only one jump operator is present with $[L(t),L(t^\prime)]=0$.

If, as a special case, one considers a pure unitary (i.e. closed system) dynamics with left generator $\mathcal L_t = -i[H(t),\cdot]$, then the right generator is the commutator with the Hamiltonian time-evolved in the \heis picture, i.e.
\begin{equation}
    \label{eq:right_gen_unitary}
    \mathcal R_t = -i\left[U^\dagger(t)H(t)U(t),\cdot\right],
\end{equation}
where $U(t)$ is the unitary evolution generated by the Hamiltonian $H(t)$.

To further illustrate the intricate relation between $\mathcal{L}_t$  and $\mathcal{R}_t$, it is convenient to consider the exponential representation of the map $\Phi_t$,
\begin{equation}
    \Phi_t = e^{G_t} ,
\end{equation}
with some map $G_t$ and the exponential is taken without time-ordering.
Clearly $G_{t=0}=0$, and let $\mathcal{M}_t =  \dot{G}_t$, i.e. $G_t = \int_0^t \mathcal{M}_\tau d\tau$; using the well known Wilcox formula \cite{Wilcox}, one has
\begin{equation}   \label{Wilcox}
    %\begin{split}
        \dot{\Phi}_t = \frac{d}{dt} e^{G_t} = \int_0^1 e^{sG_t} \mathcal{M}_t e^{(1-s)G_t}ds =  \mathcal{L}_t \circ \Phi_t ,
    %\end{split}
\end{equation}
where
\begin{equation}
    \mathcal{L}_t := \int_0^1 e^{sG_t} \mathcal{M}_t e^{-sG_t}ds . 
\end{equation}
Similarly, by changing $s \to 1-s$, one gets from \eqref{Wilcox}
\begin{equation}   \label{Wilcox-2}
    \dot{\Phi}_t = \int_0^1 e^{(1-s)G_t} \mathcal{M}_t e^{sG_t}\ ds= \Phi_t \circ \mathcal{R}_t,
\end{equation}
where
\begin{equation}
    \mathcal{R}_t := \int_0^1 e^{-sG_t} \mathcal{M}_t e^{sG_t} ds. 
\end{equation}
It is therefore clear that if $[G_t,\dot{G}_{t'}]=0$ or, equivalently, 
\begin{equation}
    \label{eq:commutativity_G}
    \left[\int_0^t \mathcal{M}_\tau d\tau,\mathcal{M}_{t^\prime}\right] = 0,
\end{equation}
then
\begin{equation}
    \mathcal{L}_t = \mathcal{M}_t = \mathcal{R}_t . 
\end{equation}
Otherwise, $\mathcal{L}_t $  and $ \mathcal{R}_t$ are different. We stress that in general $\mathcal{M}_t$ is not a generator for the dynamical map $\Phi_t$ in the sense that $\Phi_t$ is not a solution of the ME generated by $\mathcal{M}_t$. Only if commutativity \eqref{eq:commutativity_G} holds it becomes a legitimate time-local generator for $\Phi_t$ and it coincides with the left and right generators.

% ------------------------------------- Heis -------------------------------------
\subsection{\heis picture}
\label{subsec:hes_generator}
By duality, one can define the dynamical map in the Heisenberg picture $\Phi^*_t$ via
\begin{equation}
    \label{eq:duality}
    \tr\left[X\,\Phi_t[\rho]\right] = \tr\left[\Phi^*_t[X]\,\rho\right]\qquad\forall\rho,X,
\end{equation}
where $\Phi^*_t$ is the Hilbert-Schmidt adjoint of $\Phi_t$ and it is a completely positive unital (CPU) map $\Phi^*_t[\id] = \id$.
If $\Phi_t$ is unital, then $\Phi^*_t$ is also trace preserving, and viceversa.

The dual of the ME \eqref{eq:ME_left}, \eqref{eq:ME_right} reads
\begin{equation}
    \label{eq:dual_ME}
    \dot\Phi^*_t = \Phi^*_t\circ \mathcal L^*_t = \mathcal R^*_t\circ\Phi^*_t,
\end{equation}
where $\mathcal L_t^*$ and $\mathcal R_t^*$ are the Hilbert-Schmidt adjoints of $\mathcal L_t$ and $\mathcal R_t$ respectively
\begin{equation}
    \mathcal R_t^* = \dot\Phi^*_t\circ{\Phi_t^*}^{-1},\qquad
    \mathcal L_t^* = {\Phi_t^*}^{-1}\circ\dot\Phi^*_t.
\end{equation}
Notice that $\mathcal  L_t$ generates the left ME \eqref{eq:ME_left}, while its dual $\mathcal  L_t^*$ generates the right ME in the \heis picture.
The \heis left ME is instead generated by $\mathcal R^*_t$, i.e. the dual of the right generator in the \schro picture.
Equivalently, the ME \eqref{eq:dual_ME} can be written as
\begin{equation}
    \frac d{dt}X(t) = \Phi^*_t\left[\mathcal L_t^*[X(0)]\right] = \mathcal R^*_t\left[X(t)\right].
\end{equation}
Notice that in the \heis picture $\mathcal R^*_t$ generates the ME for $X(t)$, while in the \schro picture the ME for the density matrix $\rho(t)$ is generated by $\mathcal L_t$.

The generators can be written explicitly as
\begin{gather}\notag
        \mathcal R^*_t[X] = i[K(t),X]+\sum_\alpha \xi_\alpha(t)\Big[R_\alpha^\dagger(t)X R_\alpha(t)\\
        -\frac12\left\{R_\alpha^\dagger(t)R_\alpha(t),X\right\}\Big],
\end{gather}
and similarly for $\mathcal L^*_t$.
It is possible to obtain $\mathcal R^*_t$ from $\mathcal L^*_t$ via the dual of Eq.~\eqref{eq:left_to_right_gen}
\begin{equation}
    \label{eq:left_to_right_gen_dual}
    \mathcal R^*_t = \Phi_t^*\circ\mathcal L^*_t\circ{\Phi_t^*}^{-1},\quad
    \mathcal L_t^* = {\Phi_t^*}^{-1}\circ\mathcal R^*_t\circ {\Phi_t^*}.
\end{equation}
Similarly to Eqs.~\eqref{eq:Phi_time_ordered}, \eqref{eq:Phi_anti_time_ordered}, the solution of the dual ME \eqref{eq:dual_ME} can be written as either a chronological or an anti-chronological time ordered exponential 
\begin{equation}
    %\begin{split}
        \Phi^*_t = T_\rightarrow\exp\left(\int_0^td\tau\,\mathcal L^*_\tau\right)= T_\leftarrow\exp\left(\int_0^td\tau\,{\mathcal R}_\tau^*\right).
    %\end{split}
\end{equation}
Notice that when considering the dual, chronological and anti-chronological time ordering are exchanged: in the \schro picture $\mathcal L_t$ appears in a time ordered exponential, while its dual $\mathcal L_t^*$ in an anti-time ordered exponential and viceversa.
This follows from the fact that $(AB)^* = B^* A^*$ for arbitrary operators $A$ and $B$.

% ------------------------------------- Divisibility -------------------------------------
\section{Divisibility conditions}
\label{sec:divisibility}
In the \schro picture, the concept of divisibility has been widely studied and is connected to the properties of the left generator $\mathcal L_t$.
In this Section, we first briefly recall the definition of divisibility and its connection with non-Markovianity.
We then generalize this concept to the \heis picture and assign an operational interpretation to violations of divisibility also in this picture.

\subsection{Divisibility in the \schro picture}
\label{subsec:divisibility_shro}
If the dynamical map $\Phi_t$ is invertible, it is possible to define a two-parameters propagator
\begin{equation}
    \label{eq:left_divisibility}
    \Phi_{t,s}^L = \Phi_t\circ\Phi_s^{-1} = T_\leftarrow\exp\left(\int_s^td\tau\,\mathcal L_\tau\right),
\end{equation}
describing the evolution from a time $s$ to a later time $t>s$.

In a similar manner, it is possible to define a different two-parameters propagator starting from the right generator
\begin{equation}
    \label{eq:right_divisibility}
    \Phi_{t,s}^R = \Phi_s^{-1}\circ\Phi_t = T_\rightarrow\exp\left(\int_s^td\tau\,\mathcal R_\tau\right)
\end{equation}
such that the dynamical map at time $t$ can be obtained in terms of $\Phi_s$ in two ways
\begin{equation}
    \label{eq:div_2_ways_S}
    \Phi_t =\Phi^L_{t,s}\circ\Phi_s = \Phi_s\circ\Phi^R_{t,s}.
\end{equation}
We call $\Phi_{t,s}^L$ the {\it left propagator} and $\Phi_{t,s}^R$ the {\it right propagator}.
Of the two propagators, only $\Phi_{t,s}^L$ describes the time evolution of the state for intermediate times
\begin{equation}
    \label{eq:intermediate_times_S}
    \rho(t) = \Phi_{t,s}^L[\rho(s)],
\end{equation}
while for $\Phi_{t,s}^R$ one has
\begin{equation}
    \label{eq:intermediate_times_S_right_div}
    \rho(t) = \Phi_s\left[\Phi_{t,s}^R\left[\rho(0)\right]\right].
\end{equation}
This is particularly useful if one wants to computationally simulate the dynamical map: given the state $\rho(t)$ at some time $t$, the state after an infinitesimal time increment $dt$ is 
\begin{equation}
    \label{eq:infinitesimal_evol_S}
    \rho(t+dt) = \Phi_{t+dt,t}^L[\rho(t)]\approx\rho(t) + dt\,\mathcal L_t[\rho(t)].
\end{equation}
Therefore, the infinitesimal time evolution is obtained via the left generator $\mathcal L_t$.

The dynamics is said to be (C)P-divisible if the left propagator $\Phi_{t,s}^L$ is (C)PTP for all times $t\ge s$.
CP-divisibility corresponds to non-negativity of all of the rates $\gamma_\alpha(t)\ge0$ of the left generator $\mathcal L_t$, while P-divisibility is equivalent to
\cite{Kossakowski-necessary}
\begin{equation}
    \label{eq:P-div_condition}
    \sum_\alpha\gamma_\alpha(t)\abs{\braket{\varphi_\mu\vert L_\alpha(t)\vert\varphi_{\mu^\prime}}}^2\ge0
\end{equation}
for all orthonormal bases $\{\varphi_\mu\}_\mu$ and for all $\mu\ne\mu^\prime$.

Divisibility has been widely studied in the literature, and its violations have been connected to non-Markovianity of the dynamics \cite{BLP, BLP-PRA, RHP, rivas-quantum-nm, BLPV-colloquium}.
In particular, it is possible to give an operational interpretation to violations of P-divisibility as memory effects in the following way.
Suppose that Alice prepares one of two states $\rho$ or $\sigma$ randomly, each with probability $1/2$, and sends it to Bob whose task is to guess which state was prepared.
Bob's probability of correctly guessing by making a single measurement is given by \cite{Fuchs1999}
\begin{equation}
    \label{eq:P_guess_S}
    P_{\text{guess}}^{\text{s}}(\rho,\sigma) = \frac12(1+D_1(\rho,\sigma)),
\end{equation}
where $D_1$ is the trace distance \cite{Heinosaari-Ziman}
\begin{equation}
    \label{eq:TD}
    D_1(\rho,\sigma) \coloneqq \frac12\norm{\rho-\sigma}_1,\quad \norm X_1 \coloneqq \tr\abs X.
\end{equation}
Because of the contractivity of the trace distance under CPTP maps, then one has that the dynamics cannot increase $P_{\text{guess}}^{\text{s}}$ above its initial value
\begin{equation}
    P_{\text{guess}}^{\text{s}}\left(\Phi_t[\rho],\Phi_t[\sigma]\right)\le P_{\text{guess}}^{\text{s}}(\rho,\sigma).
\end{equation}
However, revivals in time of $P_{\text{guess}}^{\text{s}}$ can be present.
Notice that, since the dynamical map $\Phi_t$ is P-divisible if and only if \cite{Chruscinski2011, Wißmann-BLP-P-div, Kossakowski-necessary}
\begin{equation}
    \label{eq:contractivity_TN_S}
    \frac d{dt}\norm{\Phi_t[X]}_1\le0
\end{equation}
for all self-adjoint operators $X$, then revivals can be present only if P-divisibility is violated.
Such revivals can be interpreted as memory effects due to the presence of the environment.
It is possible to bound the revival from a time $s$ to a later time $t\ge s$ as \cite{Laine-info-bound,Megier2021,Smirne2022}
%\begin{widetext}
\begin{equation}
    \label{eq:revival_TD_bound}
    \begin{split}
        D_1\Big(\rho^1_S(t), \rho^2_S(t)&\Big) - D_1\Big(\rho^1_S(s), \rho^2_S(s)\Big) \le  D_1\Big(\rho_E^1(s),\rho_E^2(s)\Big)\\
        &+D_1\Big(\rho_{SE}^1(s),\rho^1_S(s)\otimes\rho^1_E(s)\Big) \\
        &+ D_1\Big(\rho_{SE}^2(s),\rho^2_S(s)\otimes\rho^2_E(s)\Big).
    \end{split}
\end{equation}
%\end{widetext}
In order for a revival to be present, either correlations between the system and the environment are present at time $s$ or the environmental states have become different.
It is possible to quantify the degree of non-Markovianity as \cite{BLP, BLP-PRA}
\begin{equation}
    \label{eq:BLP_S}
    \mathcal N_S(\Phi) \coloneqq \sup_{\rho,\sigma}\int\limits_{\dot\eta_S(t)>0}dt\,\dot\eta_S(t),
\end{equation}
where $\eta_S(t) \coloneqq D_1(\Phi_t[\rho],\Phi_t[\sigma])$ and the supremum is taken over all pairs of states.
If $\mathcal N_S(\Phi)>0$ then the dynamics must necessarily be non P-divisible.

Notice that the concept of divisibility can be extended also to dynamics which are not invertible \cite{Chruscinski2018, Chakraborty2021} by using the generalized inverse \cite{Ben-Israel2003} which is always guaranteed to exist even when $\Phi_t^{-1}$ does not exist.
In this work, however, we will limit ourselves to the invertible case, as is typically done with divisibility and in the framework of operational interpretation.

% ------------------------------------- Heis divisibility -------------------------------------
\subsection{Divisibility in the \heis picture}
\label{subsec:divisibility_heis}
Similarly to $\Phi_t$, also the dynamical map in the \heis picture $\Phi^*_t$ can be written in terms of the map at the intermediate time $s$ in two ways
\begin{equation}
    \label{eq:div_2_ways_H}
    \Phi^*_t = \Phi^*_s\circ \Phi^{L*}_{t,s} = \Phi^{R*}_{t,s}\circ\Phi_s^*,
\end{equation}
where $\Phi^{R*}_{t,s}$ and $\Phi^{L*}_{t,s}$ are both unital maps, reading
\begin{equation}
    \label{eq:left_divisibility_H}
    \Phi^{R*}_{t,s} = \Phi^*_t\circ{\Phi^*_s}^{-1} = T_\leftarrow\exp\left(\int_s^td\tau\,\mathcal R^*_\tau\right)
\end{equation}
and
\begin{equation}
    \Phi^{L*}_{t,s} = {\Phi_s^*}^{-1}\circ\Phi_t^* = T_\rightarrow\exp\left(\int_s^td\tau\,\mathcal L^*_\tau\right).
\end{equation}
Once again chronological and anti-chronological time orderings are swapped with respect to the \schro picture.
Accordingly, the dual of the left propagator $\Phi_{t,s}^L$ is now a right propagator $\Phi^{L*}_{t,s}$ and viceversa.
This means that the evolution from the intermediate time $s$ to the later time $t$ is now described by
\begin{equation}
    \label{eq:intermediate_times_H}
    X(t) = \Phi^{R*}_{t,s}[X(s)],
\end{equation}
and, analogously to Eq.~\eqref{eq:infinitesimal_evol_S}, the infinitesimal evolution in the \heis picture reads
\begin{equation}
    \label{eq:infinitesimal_evol_H}
    X(t+dt) = \Phi_{t+dt,t}^{R*}[X(t)]\approx X(t) + dt\,\mathcal R_t^*[X(t)].
\end{equation}
It is worth stressing that, although in the \schro picture the intermediate evolution is generated by the left generator $\mathcal L_t$, in the \heis picture it is {\it not} obtained from its dual, but from the dual of the right generator $\mathcal R_t^*$, which becomes a left generator in the \heis picture.

From Eq.~\eqref{eq:intermediate_times_H}, it is natural to define the concept of divisibility also in the \heis picture: The dynamics is said to be {\it \heis (C)P-divisible} if $\Phi^{R*}_{t,s}$ is (C)PU for all times $t\ge s$.
Because of duality, \heis divisibility is equivalent to (complete) positivity of the right propagator $\Phi_{t,s}^R$ of Eq.~\eqref{eq:right_divisibility} and not of the left propagator $\Phi_{t,s}^L$ which determines \schro divisibility.
Therefore, divisibility in the \schro and in the \heis picture are two nonequivalent concepts.
\heis CP-divisibility is equivalent to the positivity of the rates $\xi_\alpha(t)$ of the right generator $\mathcal R_t$ of Eq.~\eqref{eq:ME_right}, while for P-divisibility an analogous of Eq.~\eqref{eq:P-div_condition} holds, but using the rates $\xi_\alpha(t)$ and operators $R_\alpha(t)$ of the right generator.

Notice that, in the special case of unitary evolution in one picture, the dynamics is unitary also in the other picture, as shown in Eq.~\eqref{eq:right_gen_unitary}.
Therefore, for unitaries, divisibility is preserved when going from one picture to the other.

\subsubsection{Operational interpretation}
\label{subsec:op_interpretation_div_H}
%\note{(Sub)subsection for this part?}

\begin{figure*}
    \centering
    \includegraphics[width=0.4\linewidth]{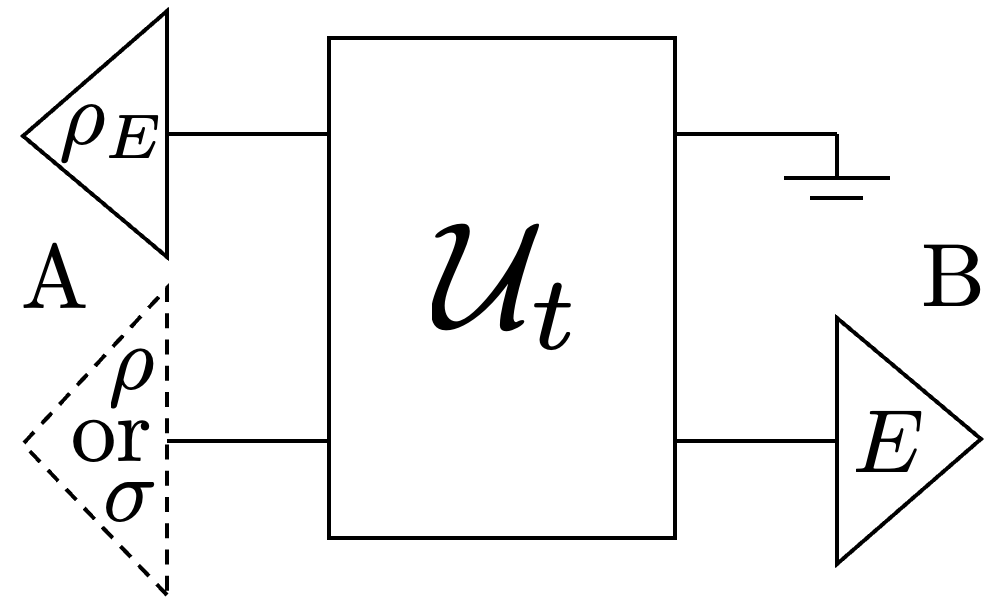}~~~~~~~~~~~
    \includegraphics[width=0.4\linewidth]{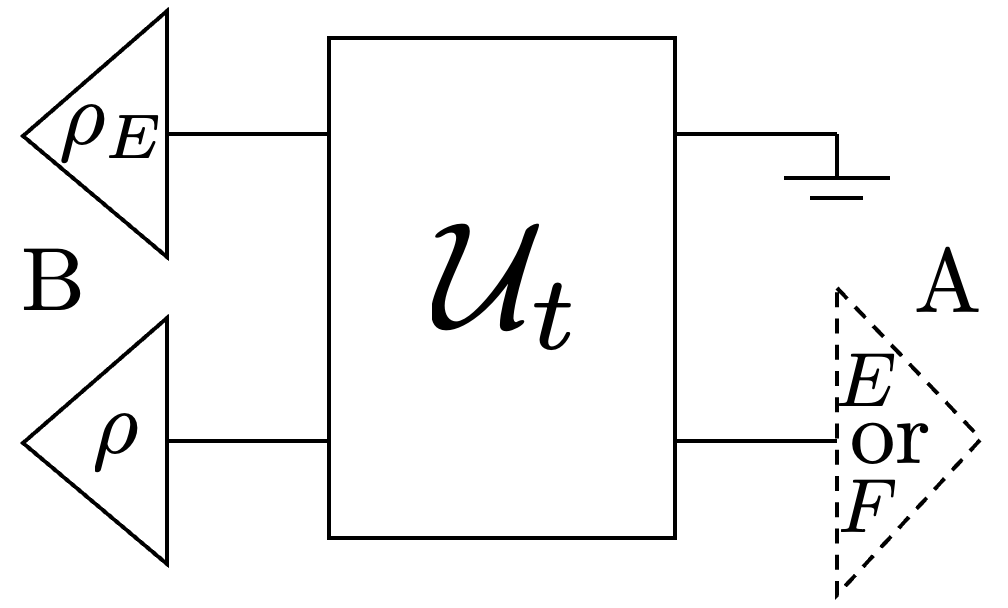}
    \caption{Left: guessing between states (\schro divisibility): Alice prepares either $\rho$ or $\sigma$ and Bob has to guess which state was prepared.
    Right: guessing between effects (\heis divisibility): Alice can measure either $E$ or $F$ and Bob has to guess which effect was measured.}
    \label{fig:P_guess_sketch}
\end{figure*}

We now proceed to show that it is possible to give an interpretation to violations of \heis P-divisibility analogous to the revivals of the guessing probability of Eq.~\eqref{eq:P_guess_S} that holds in the \schro picture.
Consider a scenario dual to the one considered for Eq.~\eqref{eq:P_guess_S}:
Alice prepares a black box performing one of two possible measurements, each with the same probability and described by effects $E$ and $F$.
Bob's task is to guess what measurement is being performed.
To do so, he can prepare a single state $\rho$ and use the black box only once.
%Alice can measure one of two possible effects $E$ or $F$, each with equal probability.
%Bob's task is to guess which effect is being measured, after Alice communicates to him the measurement outcome.
%To do so, he can prepare a single state $\rho$ and send it to Alice, that will perform a measurement.
In Appendix \ref{app:proof_P_guess_H} we show that Bob's probability of correctly guessing is
\begin{equation}
    \label{eq:P_guess_H}
    P_{\text{guess}}^{\text{e}}(E,F) = \frac12\left(1+D_\infty(E,F)\right),
\end{equation}
where $D_\infty$ is the operator distance \cite{Budini2024a}
\begin{equation}
    \label{eq:OD}
    D_\infty(E,F) \coloneqq \norm{E-F}_\infty,
\end{equation}
defined from the operator norm
\begin{equation}
    \label{eq:operator_norm}
    \begin{split}
        \norm X_\infty &\coloneqq \max\{\abs{\lambda_\alpha},\,\lambda_\alpha\text{ eigenvalue of }X\}\\
        &= \max_\psi\abs{\braket{\psi\vert X\vert\psi}}.
    \end{split}
\end{equation}
The operator norm $D_\infty$ therefore represents the bias in favor of correctly guessing which effect was measured by Alice, just like $D_1$ represents the bias in favor of guessing which state was prepared.
Since $-\id\le E-F\le\id$, the operator distance is bounded $0\le D_\infty(E,F)\le1$, and then $P_{\text{guess}}^{\text{e}}(E,F)$ is indeed a probability.
See Fig.~\ref{fig:P_guess_sketch} for a representation of the guessing probabilities in the \schro and \heis picture.

Revivals of such guessing probability can be linked to violations of divisibility.
In fact, a unital map $\Lambda$ is positive if and only if \cite{Paulsen2003}
\begin{equation}
    \label{eq:contractivity_ON_general}
    \norm{\Lambda(X)}_\infty\le\norm X_\infty
\end{equation}
for all operators $X$.
In analogy with Eq.~\eqref{eq:contractivity_TN_S}, it is possible to derive a condition for divisibility also in the \heis picture: a CPU dynamical map $\Phi^*_t$ is \heis P-divisible if and only if
\begin{equation}
    \label{eq:contractivity_ON_H}
    \frac d{dt}\norm{\Phi^*_t[X]}_\infty\le0.
\end{equation}
Therefore, the operator distance $D_\infty$ presents revivals if and only if \heis P-divisibility is violated.
It is possible to bound the revival of the distance between two effects from a time $s$ to a later time $t\ge s$ as
%\begin{widetext}
\begin{equation}
    \label{eq:revival_OD_bound}
    \begin{split}
        D_\infty\Big(X^1_S(t), X^2_S(t)&\Big) - D_\infty\Big(X^1_S(s), X^2_S(s)\Big) \le  D_\infty\Big(X_{E}^1(s),X^2_E(s)\Big)\\ 
        & + D_\infty\Big(X_{SE}^1(s),X^1_S(s)\otimes X^1_E(s)\Big)\\
        &+ D_\infty\Big(X_{SE}^2(s),X^2_S(s)\otimes X^2_E(s)\Big),
    \end{split}
\end{equation}
%\end{widetext}
where $X_{SE}^i(0) = X_S^i(0)\otimes \id_E$ with $\{X_S^i(0)\}_{i=1,2}$ two distinct effects, $X_{SE}^i(t)$ are the unitarily evolved Heisenberg picture operators, while $X_S^i(t)$ and $X_E^i(t)$ are the effects evolved according to the reduced dynamics, that is $X_S^i(t) = \tr_E\{\rho_E X_{SE}^i(t)\}$ as well as $X_E^i(t) = \tr_S\{\rho_S X_{SE}^i(t)\}$.
As discussed in more detail in Appendix \ref{app:revival_OD_bound} where the result is proven, this bound is dual to the bound of revivals of $D_1$ of Eq.~\eqref{eq:revival_TD_bound}.

Using Eqs.~\eqref{eq:P_guess_H} and \eqref{eq:contractivity_ON_H} it is thus possible to give an operational interpretation of \heis P-divisibility: if $\Phi^*_t$ is P-divisible, then the probability of guessing which effect was chosen by Alice will decrease monotonically in time -- information is lost to the environment.
If, instead, P-divisibility is violated, then there exist two effects $E$ and $F$ whose guessing probability is not monotonic.
Eq.~\eqref{eq:revival_OD_bound} allows us to interpret this revival as a backflow of information: the information initially lost in the environment can be stored as correlations (via $D_\infty(X_{SE}^i(s),X^i_S(s)\otimes X^i_E(s))\ne0$) or differences in the environments (via $D_\infty(X_{E}^1(s),X^2_E(s))\ne0$) and can later flow back into the open system at later times $t\ge s$.

In a similar manner to what was done in the \schro picture via Eq.~\eqref{eq:BLP_S}, it is possible to quantify the violation of \heis P-divisibility as
\begin{equation}
    \label{eq:BLP_H}
    \mathcal N_H(\Phi^*) \coloneqq\sup_{E,F}\int\limits_{\dot\eta_H(t)>0}dt\,\dot\eta_H(t),
\end{equation}
where $\eta_H(t) \coloneqq D_\infty(\Phi^*_t[E], \Phi^*_t[F])$ and the $\sup$ is taken over all pairs of effects $0\le E,F \le \id$.
From Eq.~\eqref{eq:contractivity_ON_H}, it holds that $\mathcal N_H(\Phi^*)>0$ if and only if the dynamics is \heis P-divisible.

Notice that revivals of the operator distance $D_\infty$ were considered in the \schro picture \cite{Budini2024a}.
However, since contractivity is guaranteed only for unital maps, revivals under completely positive but non-unital maps were interpreted as a witness of non-classicality of the dynamics \cite{Budini2023}.
Here, instead, since we are considering dynamics in the \heis picture, unitality, and thus contractivity under completely positive maps, is always guaranteed to hold.
Notice also that violations of unitality in the \schro picture are equivalent to violations of trace preservation in the \heis picture, and can be interpreted as adding biases to the measurement apparatus.

% ------------------------------------- Ph cov -------------------------------------
\subsection{Example: phase covariant dynamics}
\label{subsec:ph_cov}
As an example showing the different forms for the left and the right generators, as well as the different conditions for divisibility, let us consider the phase covariant dynamics defined by \cite{Haase-fundamental, Smirne-ultimate, Vacchini2010-notes}
\begin{equation}
    \begin{split}
        \Phi_t[\rho] =& \frac12\Big[\tr[\rho](\id+\lambda_T\sigma_z) + \lambda_z\tr[\rho\sigma_z]\sigma_z\\
        &+\lambda\left(\tr[\rho\sigma_x]\sigma_x+\tr[\rho\sigma_y]\sigma_y\right) \Big],
    \end{split}
\end{equation}
where $\lambda_T$, $\lambda$, and $\lambda_z$ are time-dependent real functions such that
\begin{equation}
    \label{eq:ph_cov_CP}
    \abs{\lambda_T}+\abs{\lambda_z}\le1,\quad 4\lambda^2+\lambda_T^2\le(1+\lambda_z)^2
\end{equation}
and $\sigma_i$ are the Pauli matrices.
The left generator reads
\begin{equation}
    \label{eq:ph_cov_left_gen}
    \begin{split}
        \mathcal L_t[\rho] = \sum_{\alpha = z,\pm}\gamma_\alpha\left[\sigma_\alpha\rho\sigma_\alpha^\dagger-\frac12\left\{\sigma_\alpha^\dagger\sigma_\alpha,\rho\right\}\right],
    \end{split}
\end{equation}
where $\sigma_+ = \ket1\bra0 = \sigma_-^\dagger$ are the raising and lowering operators, with rates
\begin{equation}
    \label{eq:ph_cov_rates_L}
    \begin{split}
        \gamma_\pm =& \frac{\pm\dot\lambda_T\lambda_z-(1+\lambda_T)\dot\lambda_z}{2\lambda_z},\\
     \gamma_z& = \frac14\left(\frac{\dot\lambda_z}{\lambda_z}-2\frac{\dot\lambda}\lambda\right).
    \end{split} 
\end{equation}
In the \schro picture, the dynamics is CP-divisible if and only if all the rates are positive $\gamma_\alpha\ge0$, while it is P-divisible if and only if \cite{Filippov-ph-cov, Teittinen2018}
\begin{equation}
    \label{eq:ph_cov_P_div}
    \gamma_\pm\ge0,\quad\text{and}\quad\gamma_z\ge-\frac12\sqrt{\gamma_+\gamma_-}.
\end{equation}

It is easy to show that the right generator $\mathcal R_t = \Phi^{-1}_t\circ\dot\Phi_t$ has the same functional form of the left generator \eqref{eq:ph_cov_left_gen} with rates
\begin{equation}
    \xi_\pm = \frac{\pm\dot\lambda_T-\dot\lambda_z}{2\lambda_z},
    \qquad
    \xi_z = \gamma_z.
    %\xi_z = \frac14\left(\frac{\dot\lambda_z}{\lambda_z}-2\frac{\dot\lambda}\lambda\right) = \gamma_z.
\end{equation}
Interestingly, the rates corresponding to $\sigma_z$ are equal ($\gamma_z=\xi_z$), while the other two rates are in general different, and therefore the left and the right generator do not coincide.
It is possible to write the right rates in terms of the left ones as
\begin{equation}
    \label{eq:ph_cov_left_to_right_rates}
    \xi_\pm = \frac{\gamma_+(\lambda_z\mp\lambda_T\pm1)+\gamma_-(\lambda_z\mp\lambda_T\mp1)}{2\lambda_z}.
\end{equation}
Notice that when the dynamics is unital ($\lambda_T=0$ or, equivalently, $\gamma_+=\gamma_-$), then the dynamics is commutative and therefore $\mathcal L_t = \mathcal R_t$.

\begin{figure}
    \centering
    \includegraphics[width=\linewidth]{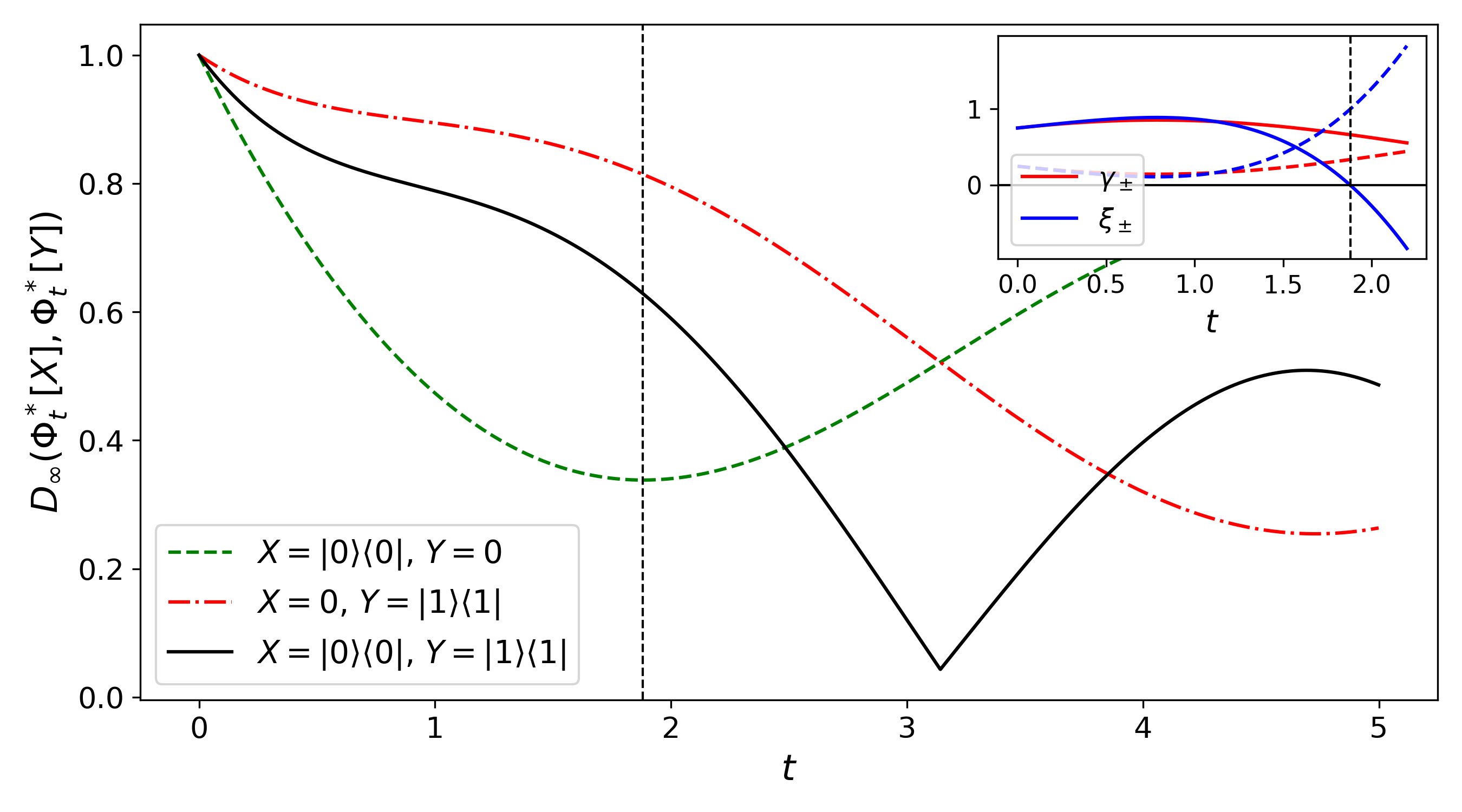}
    \caption{Phase covariant dynamics.
    Dynamics of $D_\infty(\Phi_t^*[X], \Phi^*_t[Y])$ for different initial effects $X$, $Y$.
    Inset: rates $\gamma_+$ (red solid), $\gamma_-$ (red dashed), $\xi_+$ (blue solid), and $\xi_-$ (blue dashed) corresponding to Eq.~\eqref{eq:ph_cov_counterexample}.
    At $t\approx1.9$ (vertical dashed line), $\xi_+$ becomes negative and, accordingly, $D_\infty$ behaves non-monotonically, thus witnessing violations of \heis P-divisibility.}
    \label{fig:CP_L_R_ph_cov}
\end{figure}

The conditions for \heis P- and CP-divisibility are the same as the \schro ones but with $\xi_\alpha$ instead of $\gamma_\alpha$.
However, due to the fact that $\gamma_\pm\ne\xi_\pm$, the two divisibilities are not equivalent.
In particular, it is possible to have \schro CP-divisible but \heis P-indivisible dynamics.
An explicit example is given by the choice
\begin{equation}
    \label{eq:ph_cov_counterexample}
    %\begin{split}
        \lambda_T(t) = \frac12\sin t,\quad\lambda_z(t)=e^{-t},\quad
\lambda(t)=e^{-t/2}
    %\end{split}
\end{equation}
for which $\gamma_z=\xi_z=0$, $\gamma_\pm\ge0$ but $\xi_\pm$ are not positive.
Such rates are shown in the inset of Fig.~\ref{fig:CP_L_R_ph_cov}.
The non-positivity of the right rates implies that the dynamics in the \heis picture is non P-divisible.
This fact is reflected by a non-monotonicity of $D_\infty(\Phi^*_t[E], \Phi^*_t[F])$ for some effects $E$, $F$ as a consequence of Eq.~\eqref{eq:contractivity_ON_H} and is shown in Fig.~\ref{fig:CP_L_R_ph_cov}.
The code used for obtaining the plots is available at \cite{github}.

The other way around, instead, is not possible: if the dynamics is \heis CP-divisible (and invertible) then it is also \schro CP-divisible.
In fact, the condition $\xi_\pm\ge0$ implies
\begin{equation}
    \dot\lambda_z\le\min\{\dot\lambda_T,-\dot\lambda_T\} = -\abs{\dot\lambda_T}.
\end{equation}
Substituting it in Eq.~\eqref{eq:ph_cov_rates_L} and using $\lambda_z>0$ (because of invertibility), then
\begin{equation}
    2\lambda_z\gamma_\pm\ge\abs{\dot\lambda_T}\left(1\pm\lambda_T + \lambda_z\operatorname{sgn}(\dot\lambda_T)\right),
\end{equation}
where $\operatorname{sgn}(x) = x/\abs x$ is the sign function.
Due to the CP conditions of \eqref{eq:ph_cov_CP} the term within the parentheses is always positive and therefore $\gamma_\pm\ge0$.

The fact that right divisibility implies left divisibility is specific to the particular model chosen and does not hold in general.
In Sec.~\ref{subsec:example_DO_OD} we provide explicit examples of dynamics either \heis or \schro divisible.

% ------------------------------------- Classical -------------------------------------
\subsection{Classical stochastic dynamics}
\label{subsec:classical}
Similar considerations also hold in the classical setting.
The dynamics of a $d$-site probability distribution $\mathbf{p}(t) = S(t)\,\mathbf{p}(0)$, where $\{S(t)\}_{t\geq 0}$ is a classical dynamical map represented by $d\times d$ stochastic matrix, obeys the MEs
\begin{equation}
    \label{eq:ME_classical}
    \dot S(t) = L(t) S(t)  \ , \ \ \  \dot S(t) = S(t) R(t),
\end{equation}
where $L(t)$ and $R(t)$ are, respectively, the left and right generator and are also $d\times d$ matrices.
Recall that $S(t)$, being stochastic, satisfies
\begin{equation}
    S_{ij}(t) \geq 0 \ , \ \ \ \sum_{i=1}^d S_{ij}(t) =1 ,
\end{equation}
and $S_{ij}(t=0) = \delta_{ij}$. For the dynamics to preserve the total probability, 
the corresponding right and left generators must satisfy
\begin{equation}
    \sum_{i=1}^d L_{ij}(t) = 0 \ , \ \ \ \sum_{i=1}^d R_{ij}(t) = 0 .
\end{equation}
Similarly to the quantum case, the solution can be written as
\begin{equation}
    %\begin{split}
        S(t) = T_\leftarrow\exp\left(\int_0^td\tau\,L(\tau)\right)
        = T_\rightarrow\exp\left(\int_0^td\tau\,R(\tau)\right)
    %\end{split}
\end{equation}
and the generators can be written as
\begin{equation}
    L(t) = \dot S(t)\,S(t)^{-1},\quad R(t) = S(t)^{-1}\,\dot S(t).
\end{equation}
It is possible to derive the right generator from the left one as
\begin{gather}
    R(t) = S(t)^{-1}\,L(t)\,S(t),\\L(t) = S(t)\,R(t)\,S(t)^{-1}.
\end{gather}
Also in the classical case, the two generators will coincide if and only if the dynamics is commutative, i.e. $[L(t), L(t^\prime)]=0$.

Notice that one can reformulate \eqref{eq:ME_classical} as a ME for $\mathbf{p}(t)$
\begin{equation}
    \dot{\mathbf{p}}(t) = L(t)\,\mathbf{p}(t) ,
\end{equation}
however, there is no corresponding ME for $\mathbf{p}(t)$ in terms of the right generator $R(t)$. This is why traditionally ones prefers to use the left one $L(t)$. Moving to the "Heisenberg" dual picture, one arrives at the ME for a classical observable represented by a vector $\mathbf{x} \in \mathbbm{R}^d$

\begin{equation}
    \dot{\mathbf{x}}(t) = R^\top(t)\, \mathbf{x}(t) ,
\end{equation}
where the $R^\top$ stands for the transposed matrix.
Also in this case, the ME in the dual Heisenberg pictures is defined by $R^\top(t)$ and not by $L^\top(t)$.  One finds

\begin{equation}
    (\mathbf{x}(0),\mathbf{p}(t)) =  (\mathbf{x}(t),\mathbf{p}(0)) 
\end{equation}
where $(\mathbf{a},\mathbf{b}) = \sum_k a_k b_k$, and hence

\begin{equation}
    (\mathbf{x}(0),L(t)\mathbf{p}(t)) =  (R(t)\mathbf{x}(t),\mathbf{p}(0)) .
\end{equation}

The classical map $S(t)$ is Markovian \cite{VanKampen2007}, meaning P-divisible, if it is possible to write it as
\begin{equation}
    S(t) = S^L(t,u) S(u) ,
\end{equation}
and the $S^L(t,u)$ is stochastic for all $t \geq u$. We call $S^L(t,u)$ a left propagator. 
This property is fully controlled by the left generator $L(t)$. 

\begin{proposition} The classical dynamical map $\{S(t)\}_{t\geq 0}$ is P-divisible if $L(t)$ satisfies Kolmogorov conditions, i.e.

\begin{equation}
    L_{ij}(t) \geq 0 ,  \ \ \ (i \neq j) ,
\end{equation}
for all $t \geq 0$.     
\end{proposition}
It is clear that $S^L(t,u) = S(t) S^{-1}(u)$. Similarly, one may define the right propagator $S^R(t,u) = S^{-1}(u) S(t)$ such that

\begin{equation}
    S(t) = S(u) S^R(t,u) . 
\end{equation}
One has

\begin{gather}
    S^L(t,u) = T_\leftarrow\exp\left(\int_u^td\tau\,L(\tau)\right), \\
    S^R(t,u) = T_\rightarrow\exp\left(\int_u^td\tau\,R(\tau)\right) .
\end{gather}

Notice that the positivity of the off-diagonal elements of $L_{ij}(t)$ does not imply positivity of the off-diagonal elements of $R_{ij}(t)$.  

\begin{example}
    
Let us illustrate the relation between $L(t)$ and $R(t)$ for a 2-state system. A $2\times 2$  stochastic matrix can be represented as

\begin{equation}
    S(t) = \begin{pmatrix}
        a(t) & 1-b(t)\\
        1-a(t) & b(t)
    \end{pmatrix},
\end{equation}
with $0\le a(t),b(t)\le 1$.
The generators $L(t)$ and $R(t)$ are given by
\begin{gather}
    L(t) = \dot{S}(t) S^{-1}(t) = %\frac 1{a(t)+b(t)-1}
    \begin{pmatrix}
        -\ell_1(t) & \ell_2(t)\\
        \ell_1(t) & -\ell_2(t)
    \end{pmatrix},\\
    R(t) = S(t)^{-1}\dot S(t) = %\frac 1{a(t)+b(t)-1}
    \begin{pmatrix}
       -r_1(t) & r_2(t)\\
        r_1(t) & -r_2(t)
    \end{pmatrix},
\end{gather}
where
\begin{gather}
    \label{eq:l_class}
    \ell_1(t) = \frac{ -w(t) - \dot b(t)}{a(t)+b(t)-1}, \\ 
    \label{eq:l_class_2}
    \ell_2(t) = \frac{ w(t)-\dot a(t)}{a(t)+b(t)-1} , \\
    \label{eq:r_class}
     r_1(t) = \frac{ -\dot a(t)}{a(t)+b(t)-1}, \\
    \label{eq:r_class_2}
    r_2(t) = \frac{-\dot b(t)}{a(t)+b(t)-1} ,
\end{gather}
together with $w(t) = \dot a(t) b(t) - a(t) \dot b(t)$.
Clearly the conditions for positivity of the off-diagonal elements of $L(t)$ and $R(t)$ are not equivalent. Notice that if $S(t)$ is bistochastic, implying $a(t)=b(t)$, then also $L(t) = R(t)$.

\begin{corollary}
     The classical dynamics is Schr\"odinger divisible if $\ell_k(t)\geq 0$ ($k=1,2$). Similarly, it is Heisenberg divisible if  $r_k(t)\geq 0$ ($k=1,2$).
\end{corollary}

As an example consider two scenarios

\begin{equation}    \label{ab1}
    %a(t) = \frac 12 \left(1 + e^{-t}\right) , \ \ \ b(t) = \frac 12 \left(1 + e^{- 2t}\right) ,
    a(t) = \frac {1 + e^{-t}}2, \ \ \ \ \ b(t) = \frac {1 + e^{- 2t}}2 
\end{equation}
and
\begin{equation}    \label{ab2}
    %a(t) = \frac 14 \left(3 + e^{-t}\right), \ \ b(t) = \frac 14 \left(1 + 3 e^{- 2t}\cos t\right) .
    a(t) = \frac {3 + e^{-t}}4, \ \ \ \ \  b(t) = \frac {1 + 3 e^{- 2t}\cos t}4 .
\end{equation}
The corresponding functions $\ell_k(t)$ and $r_k(t)$ are plotted in Fig. \ref{Plot-Class}. For the first scenario, the dynamics is both 
\schro and \heis divisible. However, for the second one, it is \schro divisible and \heis indivisible. In particular for any $\mathbf{x} \in \mathbbm{R}^2$ one has for both scenario

\begin{equation}   \label{dS1}
    \frac{d}{dt} \| S(t)\mathbf{x}\|_1 \leq 0 ,
\end{equation}
while 
\begin{equation}   \label{dS2}
    \frac{d}{dt} \| S^\top(t)\mathbf{x}\|_\infty \leq 0
\end{equation}
only holds for the first scenario, 
where $\|\mathbf{x}\|_\infty = \max_k |x_k|$ and $\|\mathbf{x}\|_1 = |x_1|+|x_2|$. Condition \eqref{dS2} is violated for \heis indivisible dynamics. 

\begin{figure*}
    \centering
    \includegraphics[width=0.45\linewidth]{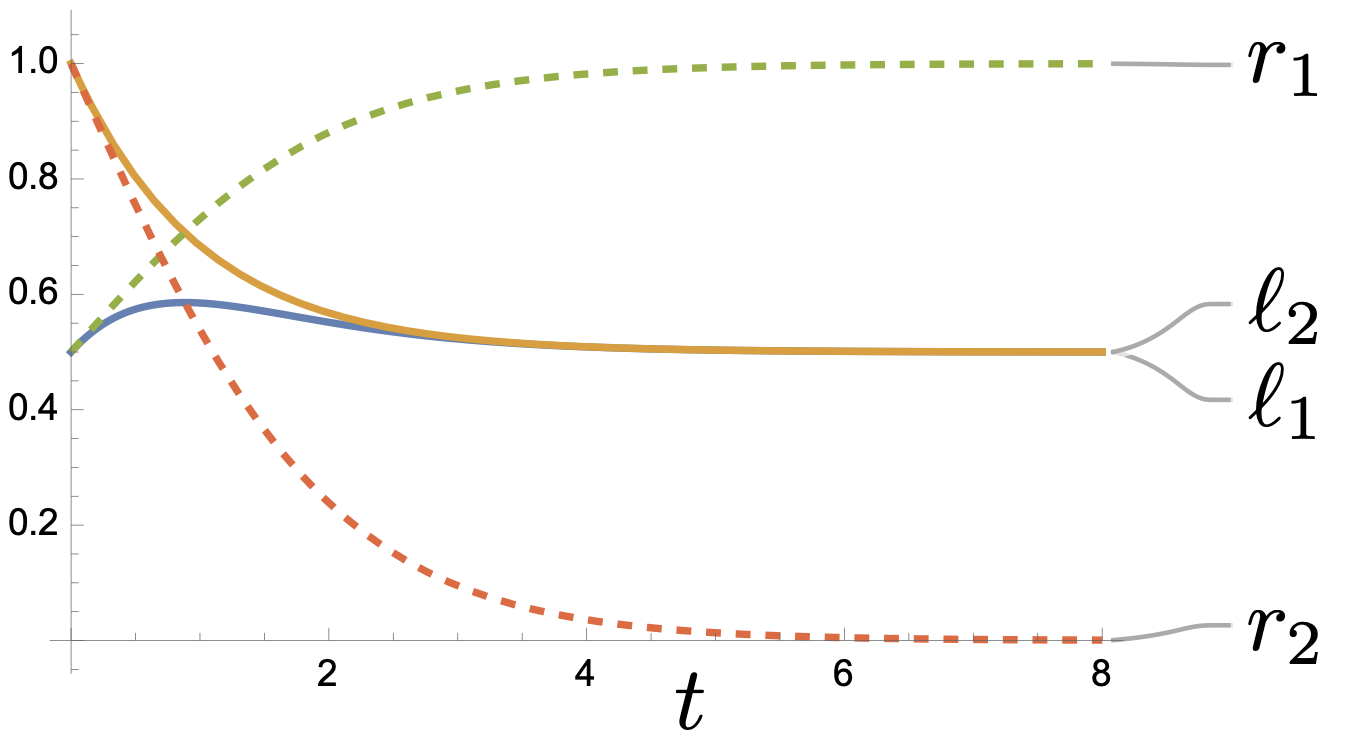}
    \includegraphics[width=0.45\linewidth]{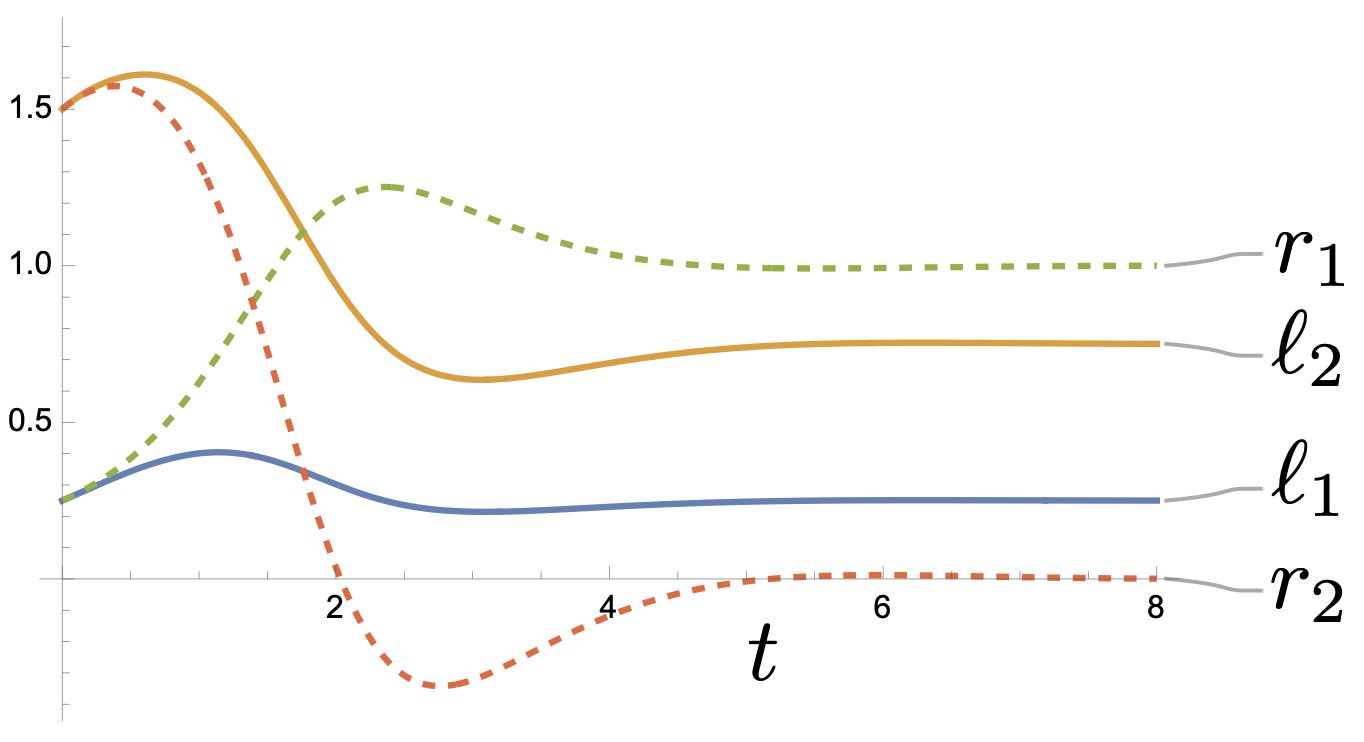}
    \caption{Off diagonal elements of $L(t)$ and $R(t)$. Solid: $\ell_{1,2}$ \eqref{eq:l_class}, \eqref{eq:l_class_2}, dashed: $r_{1,2}$ \eqref{eq:r_class}, \eqref{eq:r_class_2}. Left panel corresponds to \eqref{ab1}, right panel to \eqref{ab2}.}
    \label{Plot-Class}
\end{figure*}

Notice that, on the other hand, \heis divisibility implies \schro divisibility.
In Appendix \ref{app:classical_divs}, we show that $r_{1,2}(t)\ge0$ implies $\ell_{1,2}(t)\ge0$.
Therefore, unlike the quantum case, for classical 2-state systems, \heis divisibility is stronger than \schro divisibility.

\end{example}

% ------------------------------------- Compatibility -------------------------------------
\section{Time evolution of POVMs}
\label{sec:compatibility}

In this Section, we consider the connection between divisibility and two quantities widely used when studying POVMs: incompatibility and sharpness.
Both quantities are contractive under CPU maps, and therefore non-monotonicity is a signature of violations of \heis divisibility.

\subsection{Compatibility}
\label{subsec:compatibility}
A quantum measurement can be described by a POVM (keeping for simplicity a discrete set of measurement outcomes) $M = (M_i)_i$, which assigns 
to any quantum state a probability distribution via $\rho \mapsto (\mbox{tr}\left[\rho M_i\right])_i$.
The elements of the POVM are effects $0\le M_i\le\id$ satisfying $\sum_i M_i = \id$.
Two POVMs $M=(M_i)_i$ and $N=(N_j)_j$ are said to be {\it compatible} or {\it jointly measurable} if there exists a third POVM $G=(G_{ij})_{i,j}$ such that \cite{Busch2016}
\begin{equation}
    \label{eq:joint_measurability}
    M_i = \sum_jG_{ij},\qquad N_j=\sum_iG_{ij}.
\end{equation}
If $M$ and $N$ are projective measurements, then compatibility is equivalent to commutativity $[M_i,N_j]=0$, but in general commutativity is not necessary.
Determining whether two POVMs are compatible is generally a hard task, but it drastically simplifies if one restricts to qubits \cite{Busch2010}.

The deviation from joint measurability can be quantified via {\it incompatibility monotones}: functions $I$ on pairs of POVMs such that \cite{Heinosaari2015}
\begin{enumerate}
    \item $I(M,N) = 0$ if and only if $M$ and $N$ are compatible,
    \item $I(\Lambda[M],\Lambda[N]) \le I(M,N)$ for all CPU maps $\Lambda$.
\end{enumerate}
In the context of Bell inequalities, incompatibility is a resource: two POVMs on Alice's side are incompatible if and only if there exist a pair of POVMs on Bob's side and an entangled bipartite state such that the Clauser–Horne–Shimony–Holt (CHSH) Bell inequality \cite{Clauser1969} is violated \cite{Wolf-incompatibility-Bell}.
On the other hand, if all parties can perform the same set of measurements, any set of incompatible qubit measurements leads to the violation of some Bell inequalities in some suitable multiparticle scenario \cite{Plavala2024}.

One possible incompatibility monotone can be defined starting from the noise-deformed POVM $M^{\lambda,p}$ with elements
\begin{equation}
    M^{\lambda,p}_i \coloneqq (1-\lambda)M_i + \lambda p_i \id,
\end{equation}
which consists of mixing each POVM element with the identity, where $p=(p_i)_i$ is a probability distribution.
It is then possible to define an incompatibility monotone as the minimal mixing needed for $M$ and $N$ to be compatible
\begin{equation}
    \label{eq:incop_p}
    \begin{split}
        I_p(M,N) \coloneqq \inf\{&\lambda\ge0\,\vert\,M^{\lambda,p},\,N^{\lambda,p}\\
        &\text{ are compatible}\}.
    \end{split}
\end{equation}
Another widely used incompatibility monotone is given by
\begin{equation}
    \label{eq:incop_steering}
    \begin{split}
        I_{\text{steer}}(M,N) \coloneqq \inf\{\lambda\ge0\,\vert\,(1-\lambda)M + \lambda\Lambda_0[M]\\
        \text{ and }(1-\lambda)N + \lambda\Lambda_0[N]\text{ are compatible}\},
    \end{split}
\end{equation}
where $\Lambda_0[X] = \tr[X]\id/d$ is the completely depolarizing channel.
It is possible to interpret $I_{\text{steer}}(M,N)$ as a quantifier of incompatibility as a resource for steering \cite{Wiseman2007, Uola2020}.
%\note{Should we add the definition of steeting or at least a reference?}
In fact, it corresponds to the maximal noise that one can add to the maximally entangled state such that the resulting state is still steerable with measurements $M$ and $N$ \cite{Uola-steering}.

Non-monotonicity of incompatibility in time can be linked to violations of divisibility: because of contractivity of $I$ under PU maps, if the dynamics is \heis P-divisible then
\begin{equation}
    \label{eq:monotonicity_incop}
    \frac d{dt}I\left(\Phi_t^*(M), \Phi^*_t(N)\right) \le 0
\end{equation}
for all incompatibility monotones $I$ and for all POVMs $M$ and $N$.
If revivals of $I$ are witnessed, then the dynamics must necessarily violate \heis divisibility.
The dynamics of incompatibility in time was already studied in \cite{Addis2016, Kiukas2023}, however, they considered commutative dynamics, for which \schro and \heis divisibility coincide.
In Sec.~\ref{subsec:example_DO_OD}, instead, we will provide explicit examples in which revivals of the resources quantified by $I_p$ and $I_{\text{steer}}$ are present even if the dynamics is Markovian in the \schro picture.
Similar results were derived in \cite{Kiukas-steering-EPR}, in which it was shown that steering can be enhanced in noisy \schro divisible qubit dynamics by applying a time-dependent control Hamiltonian.
There, however, the revival of steering was not connected to violations of \heis divisibility and, unlike in \cite{Addis2016, Kiukas2023}, the dynamics is not commutative due to the time-dependence of the Hamiltonian.

\subsection{Sharpness}
\label{subsec:sharpness}

Another important quantity characterizing POVMs is that of {\it sharpness}, quantifying how close a POVM is to an ideal projective measurement.
Similarly to incompatibility, sharpness can also be defined in non-unique ways.
Given an effect $E$, a definition of the sharpness of the binary POVM $(E, \id-E)$ is \cite{Busch2009}
\begin{equation}
    \label{eq:sharpness}
    \Sigma(E) \coloneqq \norm E_\infty + \norm{\id-E}_\infty - 1.
\end{equation}
Sharpness is contractive under PU maps, which follows from contractivity of the operator norm.
Therefore, similarly to incompatibility, \heis P-divisibility implies
\begin{equation}
    \label{eq:monotonicity_sharpness}
    \frac d{dt} \Sigma\left( \Phi^*_t[E]\right)\le0.
\end{equation}
Notice that this condition is not equivalent to contractivity of the operator norm \eqref{eq:contractivity_ON_H} since it involves both $E$ and $\id-E$.
For instance, the phase covariant example of Eq.~\eqref{eq:ph_cov_counterexample} presents revivals in the operator norm but not of sharpness: revivals of $\norm E_\infty$ are compensated by decreases of $\norm{\id-E}_\infty$.
Nevertheless, in Sec.~\ref{subsec:example_DO_OD} we will show an explicit example with a complete revival of sharpness for a dynamics \schro divisible but \heis indivisible.
Again, it is possible to have revivals of resources in the \heis picture without having non-Markovianity in the \schro picture.

% ------------------------------------- Qubit -------------------------------------
\section{Explicit conditions for qubit}
\label{sec:qubit}
\renewcommand{\vec}{\mathbf}
In this Section, we derive explicit conditions for P-divisibility for maps acting on qubits.
The state of a qubit can be described via a three dimensional vector $\vec r=(x,y,z)^\top$, $\norm{\vec r}\le1$ where $\norm\cdot$ is the Euclidean norm, of the Bloch sphere $\mathcal S$  such that
\begin{equation}
    \rho = \frac12\left(\id + \vec r\cdot\vec\sigma\right)\leftrightarrow\vec r,
\end{equation}
where $\vec\sigma=(\sigma_x,\sigma_y,\sigma_z)^\top$.
As a consequence, any CPTP map $\Phi$ 
%(in the following, we omit time dependence) 
can be written as a $4\times4$ matrix
\begin{equation}
    \label{eq:Bloch_isomorphism_map_S}
    \Phi\leftrightarrow M_\Phi = \begin{pmatrix}
        1 & \vec 0^\top\\
        \vec v &\Lambda
    \end{pmatrix},
\end{equation}
where $\Lambda$ is a $3\times 3$ real matrix and $\vec v$ a 3-dimensional real vector, acting on the 4-dimensional Bloch vector $\vec r_4 = (1,\vec r)^\top$ so that the corresponding transformation of the Bloch sphere is the affine transformation
\begin{equation}
    \Phi[\rho]\leftrightarrow\Lambda\vec r+\vec v.
\end{equation}
The map $\Phi$ is positive if and only if it maps $\mathcal S$ into itself, while conditions for its complete positivity in terms of the Bloch representation have been derived in \cite{Ruskai2002}.

For self-adjoint operators, the Bloch representation consists of the 4-dimensional real vector $\vec a_4 = (a_0,a_x,a_y,a_z)^\top = (a_0,\vec\alpha)^\top$ such that
\begin{equation}
        A = \frac12\left(a_0\id + \vec\alpha\cdot\vec\sigma\right).
\end{equation}
Furthermore, $A$ is an effect if and only if \cite{Heinosaari-Ziman}
\begin{equation}
    0\le a_0\le2, \quad \norm{\vec\alpha}\le\min\left\{a_0,2-a_0\right\}.
\end{equation}

It is easy to show that
\begin{equation}
    \tr[A\rho] = \vec a_4^\top\cdot\vec r_4 = a_0 + \vec\alpha\cdot\vec r,
\end{equation}
so that the matrix representation of the dual map $\Phi^*$ is given by
\begin{equation}
    \begin{split}
        \tr[A\Phi[\rho]] &= \vec a_4^\top\cdot M_\Phi \vec r_4 = (M_\Phi^\top\vec a_4)^\top\cdot\vec r_4 \\
        &= \tr[\Phi^*[A]\rho] = (M_{\Phi^*}\vec a_4)^\top\cdot\vec r_4
    \end{split}
\end{equation}
and therefore
\begin{equation}
    M_{\Phi^*} = M_\Phi^\top.
\end{equation}

Using (we omit time dependence for compactness)
\begin{gather}
    M_{\dot\Phi} = \frac d{dt}M_\Phi,\quad M_{\Phi_1\Phi_2} = M_{\Phi_1}M_{\Phi_2},\\M_{\Phi^{-1}} = M_\Phi^{-1},%=\begin{pmatrix}1&\vec 0^\top\\-\Lambda^{-1}\vec v&\Lambda^{-1}\end{pmatrix},
\end{gather}
it is straightforward to obtain the matrix representation of the left and right generators
\begin{gather}
    M_{\mathcal L} = \begin{pmatrix}
        0&\vec 0^\top\\
        \dot{\vec v}-\dot\Lambda\Lambda^{-1}\vec v & \dot\Lambda\Lambda^{-1}
    \end{pmatrix},\\
    M_{\mathcal R} = \begin{pmatrix}
        0&\vec 0^\top\\
        \Lambda^{-1}\dot{\vec v} & \Lambda^{-1}\dot\Lambda
    \end{pmatrix},
\end{gather}
and therefore the Heisenberg generators are
\begin{gather}
    M_{\mathcal L^*} = \begin{pmatrix}
        0&(\dot{\vec v}-\dot\Lambda\Lambda^{-1}\vec v)^\top\\
        \vec 0 & \Lambda^{-\top}\dot\Lambda^\top
    \end{pmatrix},\\
    M_{\mathcal R^*} = \begin{pmatrix}
        0&(\Lambda^{-1}\dot{\vec v})^\top\\
        \vec 0 & \dot\Lambda^\top\Lambda^{-\top}
    \end{pmatrix},
\end{gather}
where $\Lambda^{-\top} = (\Lambda^{-1})^\top = (\Lambda^\top)^{-1}$.

% ------------------------------------- Constraints div qubit -------------------------------------
\subsection{Analytic constraints on divisibility}
\label{subsec:constraints_qubit}
In the \schro picture, the dynamics is P-divisible if and only if $\Phi_{t+dt,t}$ maps the Bloch ball into itself.
Using the Bloch matrix representation for $\Phi_{t+dt,t}^L$ one has
\begin{equation}
     M_{\Phi_{t+dt,t}^L} = 
    \begin{pmatrix}
        1&\vec 0^\top\\
        dt\left(\dot{\vec v}_t-\dot\Lambda_t\Lambda^{-1}_t\vec v_t\right) & \id + dt\,\dot\Lambda_t\Lambda^{-1}_t
    \end{pmatrix}
\end{equation}
and therefore P-divisibility is equivalent to $M_{\Phi_{t+dt,t}^L}$ mapping $\mathcal S$ into itself
\begin{equation}
    \label{eq:P-div-Bloch-Shro}
    \norm{\left(\id + dt\,\dot\Lambda_t\Lambda^{-1}_t\right)\vec r + dt\left(\dot{\vec v}_t-\dot\Lambda_t\Lambda^{-1}_t\vec v_t\right)}\le1
\end{equation}
for all $\vec r\in\mathbf S$.

In the \heis picture, P-divisibility corresponds to $\Phi_{t+dt,t}^{R*}$ being PU.
Notice that positivity of $\Phi_{t+dt,t}^{R*}$ is equivalent to $\Phi_{t+dt,t}^R$ being PTP.
On the Bloch sphere, the corresponding matrix reads
\begin{equation}
    M_{\Phi_{t+dt,t}^R} = 
    \begin{pmatrix}
        1&\vec 0^\top\\
        dt\,\Lambda^{-1}_t\dot{\vec v}_t & \id+dt\,\Lambda^{-1}_t\dot\Lambda_t
    \end{pmatrix}
\end{equation}
and therefore \heis P-divisibility corresponds to 
\begin{equation}
    \label{eq:P-div-Bloch-Heis}
    \norm{\left(\id+dt\,\Lambda^{-1}_t\dot\Lambda_t\right)\vec r + dt\,\Lambda^{-1}_t\dot{\vec v}_t}\le1
\end{equation}
for all $\vec r\in\mathbf S$.

\begin{figure*}
    \centering
    \includegraphics[width=\linewidth]{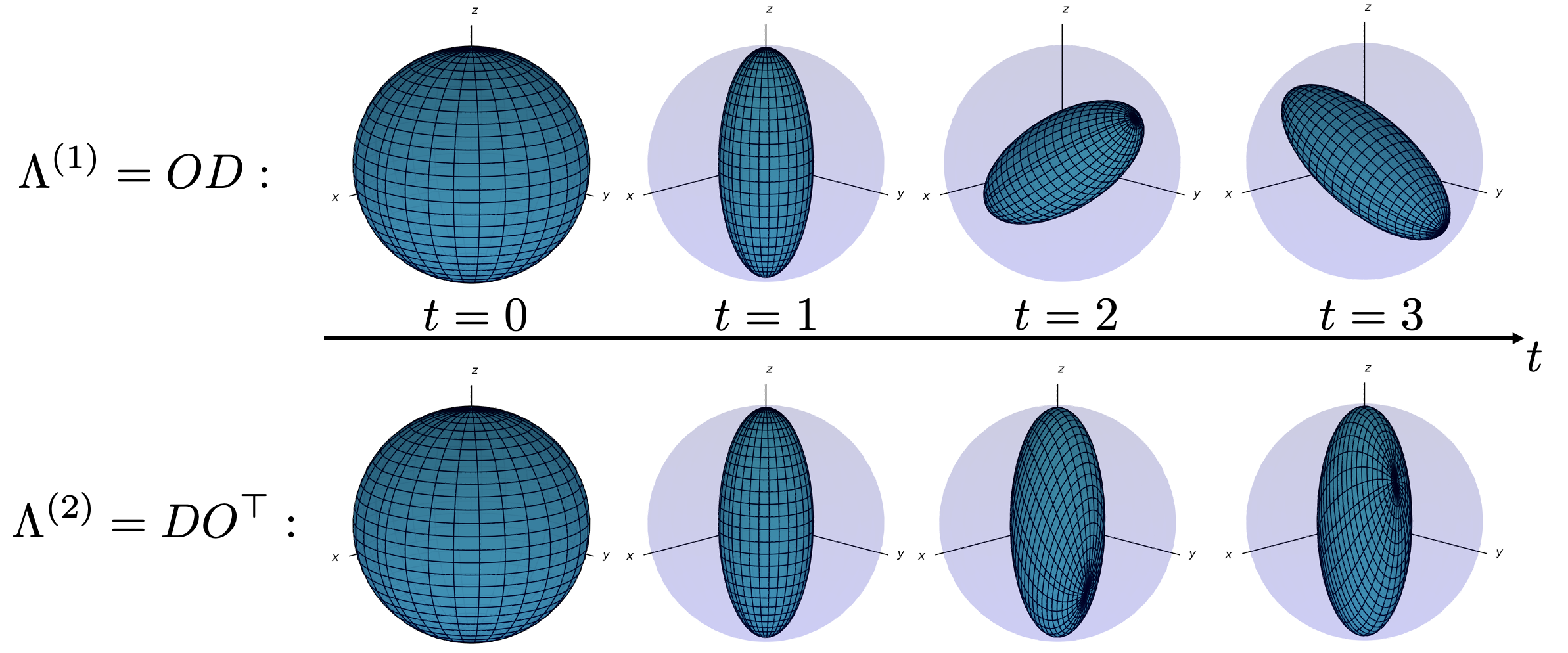}
    \caption{Representation of the time evolution of the Bloch sphere under the maps $\Lambda^{(1,2)}$ of Eq.~\eqref{eq:OD_and_DO}, in units $t_1=1$.
    For $t\le1$, $\Lambda^{(1)}=\Lambda^{(2)}$.
    At later times, $\Lambda^{(1)}$ is just an orthogonal rotation of the ellipsoid giving the evolution of the Bloch sphere, while $\Lambda^{(2)}$ does not change the image of the ellipsoid, 
    with the image of the pure states moving along its surface.}
    \label{fig:dephasing-sketch}
\end{figure*}

\subsubsection{Unital maps}
In the remaining, we will focus on unital maps in the \schro picture (or, equivalently, on trace preserving maps in the \heis picture).
This condition is equivalent to $\vec v_t = \dot{\vec v}_t = 0$ and therefore the Bloch matrices for both $\Phi_{t+dt,t}^L$ and $\Phi_{t+dt,t}^R$ are in a block-diagonal form.

The condition for P-divisibility in the \schro picture \eqref{eq:P-div-Bloch-Shro} takes the simpler form
\begin{equation}
    %\begin{split}
        \sup_{\vec r\in\mathcal S}\norm{\left(\id + dt\,\dot\Lambda_t \Lambda^{-1}_t\right)\vec r}^2
        = \norm{\id + dt\,\dot\Lambda_t \Lambda^{-1}_t}_\infty^2 \le 1,
    %\end{split}
\end{equation}
while \heis divisibility \eqref{eq:P-div-Bloch-Heis} reads
\begin{equation}
    \norm{\id+dt\,\Lambda^{-1}_t\dot\Lambda_t}_\infty^2\le1.
\end{equation}
These conditions can be simplified further by noticing that for an arbitrary matrix $X$, at first order in $dt$ and assuming $\norm{\vec r}=1$, it holds
\begin{equation}
    \norm{\left(\id + dt\,X\right)\vec r}^2
    %= \braket{\vec r,\left(\id + dt\,X^\top\right)\left(\id + dt\,X\right)\vec r}
    = 1 + dt\braket{\vec r,\left(X+X^\top\right)\vec r}
\end{equation}
and therefore $\norm{\id + dt\,X}_\infty\le1$ if and only if $X+X^\top\le0$.
Thus, \schro divisibility is equivalent to \cite{Benatti2024}
\begin{equation}
    \label{eq:P_div_S_negativity}
    X_S\coloneqq\dot\Lambda_t\Lambda^{-1}_t + \Lambda^{-\top}_t\dot\Lambda^\top\le0,
\end{equation}
while for \heis
\begin{equation}
    \label{eq:P_div_H_negativity}
    X_H\coloneqq\Lambda^{-1}_t\dot\Lambda_t + \dot\Lambda^\top\Lambda^{-\top}_t\le0.
\end{equation}

The map $\Lambda$ can always be written as \cite{King2001, Ruskai2002}
\begin{equation}
    \Lambda = O_1 D O_2^\top,
\end{equation}
where $D = \operatorname{diag}\{\lambda_1,\lambda_2,\lambda_3\}$, and $O_{1,2}$ are orthogonal matrices with $\det O_{1,2}=1$.
To compute the P-divisibility conditions, we need to compute
\begin{equation}
    \dot\Lambda = \dot O_1DO_2^\top+O_1\dot DO_2^\top + O_1 D \dot O_2^\top.
\end{equation}
The derivative of the diagonal term is easily written as $\dot D = \operatorname{diag}\{\dot\lambda_1,\dot\lambda_2,\dot\lambda_3\}$.
The orthogonal terms, instead, can always be written as
\begin{equation}
    O_i = e^{A_i},\qquad A_i^\top = -A_i.
\end{equation}
Using the Baker–Campbell–Hausdorff formula, it is possible to compute $O_i(t+dt)$ at first order in $dt$ as
\begin{equation}
    \begin{split}
        e^{A_i(t+dt)} =& e^{A_i+dt\dot A_i} = e^{A_i}e^{dt\dot A_i}e^{C_i\,dt}\\
        =&e^{A_i}\left(I+dt\dot A_i\right)\left(I+C_i\,dt\right)\\
        =& e^{A_i}\left[I+dt\left(\dot A_i+C_i\right)\right]\\
        \eqqcolon& e^{A_i}[I+dt\,B_i], 
    \end{split}
\end{equation}
where $C_i$ is an expression involving nested commutators of $A_i$ and $\dot A_i$, and therefore  $C_i^\top = -C_i$.
Hence, it holds that
%containing nested commutators $[X_1,[X_2,\ldots,[X_{n-1},X_n]\ldots]]$ and $[\ldots[[X_1,X_2], X_3],\ldots,X_n]$, where $X_j$ is either $A_i$ or $\dot A_i$, and therefore $X_j^\top = -X_j$ which implies $C_i^\top = -C_i$.
\begin{equation}
    \dot O_i(t) = \frac{O_i(t+dt)-O_i(t)}{dt} = e^{A_i(t)} B_i(t),
\end{equation}
with $B_i^\top(t) = -B_i(t)$.
Therefore, using $\Lambda^{-1} = O_2D^{-1}O_1^\top$ and $[D^{-1},\dot D]=0$, the conditions for P-divisibility in the \schro and \heis picture of Eqs.~\eqref{eq:P_div_S_negativity}, \eqref{eq:P_div_H_negativity} become
\begin{gather}
    \label{eq:X_S_unital_generic}
    \begin{split}
        X_S = e^{A_1}\Big[&2 \dot D D^{-1} - DB_2D^{-1} \\
        &+ D^{-1}B_2D\Big]e^{-A_1}\le0,
    \end{split}
    \\
    \label{eq:X_H_unital_generic}
    \begin{split}
        X_H = e^{-A_2}\Big[&2 \dot D D^{-1} - D^{-1}B_1D \\
        &+ DB_1D^{-1}\Big]e^{A_2}\le0.
    \end{split}
\end{gather}
The first term $2 \dot DD^{-1}$ is common to both divisibility conditions and its negativity is equivalent to monotonic contraction of the Bloch sphere $\dot\lambda_i\le0$.
The orthogonal transformations, instead, play a different and non-trivial role in the two pictures.
In the \schro picture, $O_1=e^{A_1}$ does not play any role; $O_2$, on the other hand, can change the positivity of $X_S$ whenever $\dot O_2\ne0$.
The situation is analogous in the \heis picture, upon swapping $O_1$ and $O_2$.
Notice that we have considered unital maps only for the sake of simplicity: numerical evidence suggests that it is possible to find maps that are P-divisible only in one picture also when unitality is broken.

\begin{figure*}
    \centering
    \includegraphics[width=\linewidth]{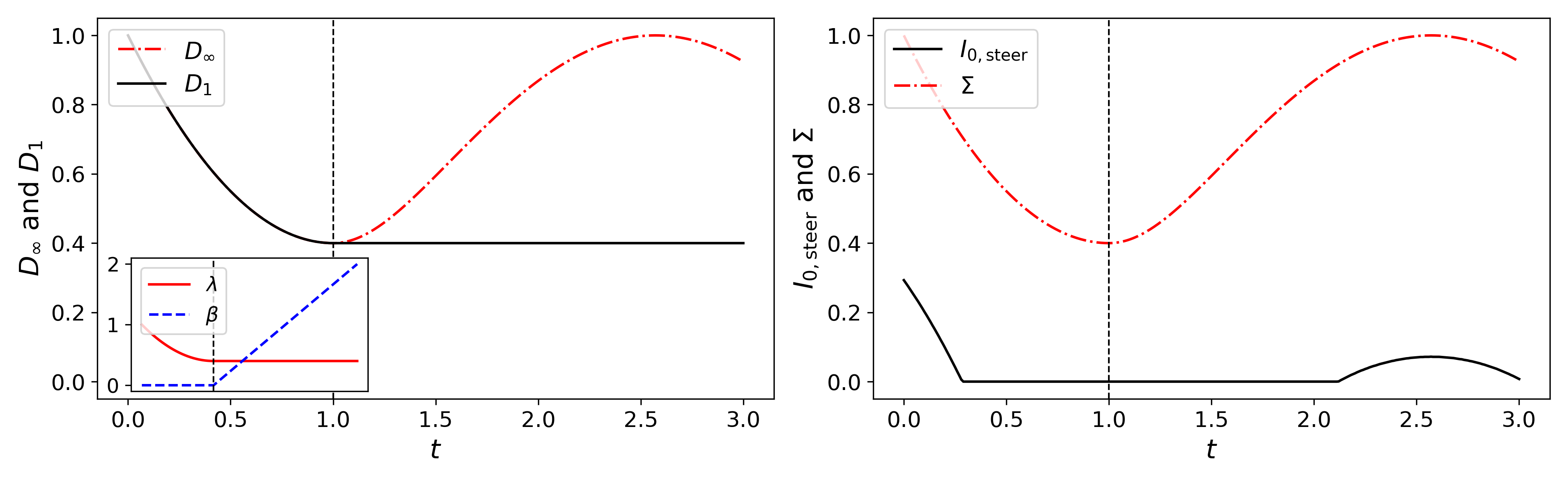}
    \caption{Dynamics of $\Phi_t$ corresponding to $\Lambda^{(1)}$ of Eq.~\eqref{eq:OD_and_DO}.
    Left panel: time evolution of $D_\infty\left(\Phi^*_t\left[\ketbra{+_y}\right], \Phi^*_t\left[\ketbra{-_y}\right]\right)$ and of $D_1\left(\Phi_t\left[\ketbra{+_y}\right], \Phi_t\left[\ketbra{-_y}\right]\right)$, where $\ket{\pm_y}$ are the eigenstates of $\sigma_y$. Since the dynamics is \schro CP-divisible but \heis P-indivisible, $D_1$ is monotonic while $D_\infty$ is not.
    Inset: dynamical functions $\lambda(t)$ and $\beta(t)$ of Eqs.~\eqref{eq:D_dephasing}, \eqref{eq:O_dephasing}.
    Right panel: dynamics of incompatibility $I_0\left(\Phi^*_t\left[\ketbra{+_y}\right], \Phi^*_t\left[\ketbra{+_x}\right]\right)$ (for this particular dynamics, $I_0=I_{\text{steer}}$) and sharpness $\Sigma\left(\Phi^*_t\left[\ketbra{+_y}\right]\right)$ of Eqs.~\eqref{eq:incop_p}, \eqref{eq:incop_steering}, and \eqref{eq:sharpness}.
    $\ket{+_x}$ is the eigenstate of $\sigma_x$.
    Notice that the dynamics presents a full revival in sharpness and only a partial revival in incompatibility.
    The vertical lines represent the time at which \heis P-divisibility is broken.}
    \label{fig:dephasing}
\end{figure*}

% ------------------------------------- Example D O - O D -------------------------------------
\subsection{Example: dephasing and orthogonal transformation}
\label{subsec:example_DO_OD}
We now proceed to derive explicit examples of dynamics either only divisible in the \schro picture or only in the \heis picture.
Suppose that, at some given time $t>0$, the diagonal term is such that $\dot D = 0$.
Then the P-divisibility conditions of Eqs.~\eqref{eq:X_S_unital_generic}, \eqref{eq:X_H_unital_generic} become
\begin{gather}
    X_S = D^{-1}B_2D - DB_2D^{-1}\le0,\\
    X_H = DB_1D^{-1} - D^{-1}B_1D\le0.
\end{gather}
If $D=\lambda\id$ (or if $\dot O_1=\dot O_2=0$), then both conditions are trivially satisfied, since $X_S=X_H=0$.
In the general case of non-isotropic shrinking and non-constant orthogonal transformations, instead, such conditions are never satisfied: it always holds that $\tr X_S=\tr X_H=0$, and therefore, if $X_{S,H}\ne0$, they must have both positive and negative eigenvalues, thus violating divisibility.
For instance, if $B_2=0$ ($\dot O_2=0$) and $B_1\ne0$ ($\dot O_1\ne0$), then the dynamics is not P-divisible in the \heis picture, while it is just a unitary rotation in the \schro picture, and therefore it is CP-divisible.
If, instead, $B_2\ne0$ ($\dot O_2\ne0$) and $B_1=0$ ($\dot O_1=0$), it is the other way around.

Let us now consider a concrete example.
More explicitly, we consider two dynamical maps $\Phi^{(1,2)}$ such that their Bloch matrix representations $\Lambda^{(1,2)}$ have only one of the orthogonal transformations non trivial, namely
\begin{equation}
    \label{eq:OD_and_DO}
    \Lambda^{(1)}=O D,\qquad \Lambda^{(2)} = D O^\top.
\end{equation}
In order to emphasize the different roles of $D$ and $O$ in the two pictures, we consider a scenario in which for $t<t_1$ only the diagonal term plays a role ($O(t<t_1)=\id$), while for $t\ge t_1$ only the orthogonal term does ($D(t\ge t_1)=D(t_1)$ or, equivalently, $\dot D(t\ge t_1)=0$).
For the sake of the example, we fix $D$ as dephasing
\begin{equation}
    \label{eq:D_dephasing}
    D(t) = \operatorname{diag}\{\lambda(t),\lambda(t),1\},
\end{equation}
with $\lambda(0)=1$ and $\dot\lambda(t)\le0$ to ensure continuity and CP-divisibility in both pictures for $t<t_1$, and $\lambda(t\ge t_1) = \lambda(t_1)$ to have $\dot D(t\ge t_1)=0$.
The orthogonal transformation, instead, reads
\begin{equation}
    \label{eq:O_dephasing}
    O(t) = \exp[\beta(t)\,\Theta(t-t_1)\,L_x],
\end{equation}
where $\Theta$ is the Heaviside theta function and $L_x$ is the generator of the rotations in the $x$ direction.
A pictorial representation of the dynamics of the Bloch sphere under the maps $\Lambda^{(1,2)}$ of Eq.~\eqref{eq:OD_and_DO} is presented in Fig.~\ref{fig:dephasing-sketch}, while the explicit form of $\lambda(t)$ and $\beta(t)$ is shown in the inset of Fig.~\ref{fig:dephasing}.
Animations representing the time evolution of the Bloch sphere can be found at \cite{github}.

For $t<t_1$, $\Lambda^{(1)}(t<t_1) = \Lambda^{(2)}(t<t_1)$ and the dynamics is CP-divisible in both pictures.
For $t\ge t_1$, let us focus on $\Lambda^{(1)}=O D$.
In the \schro picture, it is straightforward to verify that $X_S^{(1)}(t\ge t_1)=0$ and the dynamics is simply a rotation of the shrinked Bloch sphere
\begin{equation}
    \label{eq:L_t>t1_dephasing}
    \mathcal L_{t\ge t_1}^{(1)}[\rho] = -i[H^{(1)}(t),\rho].
\end{equation}
In the \heis picture, instead, the dynamics is not P-divisible, as it can be seen by explicitly computing the divisibility condition of Eq.~\eqref{eq:X_H_unital_generic}, which reads
\begin{equation}
    \begin{split}
        X_H^{(1)}(t&\ge t_1) = \dot\beta(t)\Big[D(t_1)L_xD^{-1}(t_1)\\
        &-D(t_1)^{-1}L_xD(t_1)\Big]\ne 0.
    \end{split}
\end{equation}
But, since $\tr X_H^{(1)}(t\ge t_1)=0$, then necessarily $X_H^{(1)}(t\ge t_1)\not\le0$ and therefore the dynamics is not \heis P-divisible for $t\ge t_1$.
This fact can be seen by looking at the behavior in time of the trace distance $D_1$ and of the operator distance $D_\infty$ (see left panel of Fig.~\ref{fig:dephasing}): the former is monotonically decreasing, corresponding \schro P-divisibility, while the latter is non-monotonic, and therefore \heis P-divisibility is violated.
Therefore, it holds that $\mathcal N_S(\Phi^{(1)})=0$, while $\mathcal N_H(\Phi^{(1)*})>0$.
In a similar manner, both incompatibility \eqref{eq:incop_p}, \eqref{eq:incop_steering} and sharpness \eqref{eq:sharpness} quantifiers are non-monotonic (see right panel of Fig.~\ref{fig:dephasing}), 
showing that they witness \heis divisibility violations.

On the other hand, if instead of $\Lambda^{(1)}$ we consider $\Lambda^{(2)} = D O^\top$, then the opposite scenario arises.
In the \heis picture, the dynamics is simply a unitary rotation and, correspondingly, $X_H^{(2)}(t\ge t_1)=0$, implying P-divisibility.
This time, instead, \schro P-divisibility is violated, as it can be seen by computing
\begin{equation}
    X_S^{(2)}(t\ge t_1) = -X_H^{(1)}(t\ge t_1)\not\le0.
\end{equation}
Accordingly, it holds that $\mathcal N_S(\Phi^{(2)})>0$, while $\mathcal N_H(\Phi^{(2)*})=0$: now $D_\infty$ is monotonic, while $D_1$ presents revivals.
Because of \heis divisibility, both incompatibility and sharpness are also monotonic.
Notice that the dynamics is non-Markovian in the \schro picture even if the image of the Bloch sphere remains constant: 
there is no expansion of the Bloch sphere, but non-Markovianity traces back to a non-trivial motion of its surface.

Therefore, considering situations for which $\dot D=0$, it is possible to have dynamics that are not divisible either in the \schro or in the \heis picture by simply adding a non-constant orthogonal transformation either to the right or to the left of $D$, respectively.
By means of this simple example, we have shown that the two concepts of divisibility are not equivalent and it can be violated in either only one picture or in both at the same time.

Moreover, the example allows one to highlight the crucial role of the orthogonal transformations in determining non-divisibility.
Indeed, if only $D$ is present ($O(t)=\id$ also for $t\ge t_1$), then $\Lambda^{(1)}=\Lambda^{(2)}=D$ and the dynamics is divisible in both pictures.
The addition of a nontrivial orthogonal transformation $O$ either to the left or to the right of $D$ allows the dynamics in one picture to be non-Markovian, although corresponding only to a unitary rotation in the other picture.
This fact is possible because $O$ breaks the commutativity \eqref{eq:commutativity}.
Therefore, even if the condition of \schro P-divisibility \eqref{eq:P-div_condition} does not depend on the Hamiltonian part of the generator, such non-trivial Hamiltonian in the \schro picture might alter the divisibility in the \heis picture.

Notice that in this example the dynamics is not smooth at $t=t_1$, however the definition of $O$ and $D$ can be changed to have smooth versions of the $\Theta$ function and the behavior in time is left unchanged.
Similarly, in order to emphasize the different roles of the diagonal and of the orthogonal parts of the dynamics, we have considered the simplified scenario in which only either $D$ or $O$ is non-constant at each time.
However, if we omit the theta function $\Theta$ altogether in the definition of $O$ in Eq.~\eqref{eq:O_dephasing}, the qualitative behavior of the two time evolutions is the same: $\Lambda^{(1)}$ presents violations of P-divisibility only in the \heis picture, while $\Lambda^{(2)}$ only in the \schro picture.

% ------------------------------------- Conclusions -------------------------------------
\section{Conclusions}
\label{sec:conclusions}
In this work, we have introduced the concept of divisibility for dynamical evolutions in the \heis picture.
While the \schro and \heis pictures are indeed physically equivalent in terms of observable predictions, we have shown that divisibility in one picture does not imply divisibility in the other.
This happens because the time-ordered \heis picture propagator $\Phi^{R*}_{t,s}$ is not generated by the dual of the \schro generator, and therefore its (complete) positivity is not equivalent to the (complete) positivity of the \schro propagator $\Phi^{L}_{t,s}$.
We have illustrated this fact with concrete examples whose evolution is divisible only in one picture.
This inequivalence implies that common quantifiers of non-Markovianity may miss relevant features of the dynamics if it is considered only in the \schro picture.

We have also provided an operational interpretation for violations of P-divisibility in the \heis picture which is dual to the interpretation in the \schro picture: just like violations in the \schro picture imply a non-monotonic probability of guessing which of two states is prepared, violations in the \heis picture imply non-monotonicity 
in the guessing probability of which of two effects is measured.
According to this interpretation, we have also provided a measure of the total violation of divisibility in the \heis picture analogous to the measure of non-Markovianity in the \schro picture.
Such violations of divisibility can also be connected to a non-monotonic behavior of compatibility or sharpness of POVMs.

Our work shows that memory effects can be present in the dynamics even without corresponding to violations of divisibility in the \schro picture, since divisibility can be violated only in the \heis picture.
These memory effect can be seen from the non-monotonic guessing probability between effects.
Our results align with the findings of \cite{Milz2019}, in which it was shown that CP divisibility in the \schro picture does not imply Markovianity.
This also suggests that CP divisibility, either in the \schro or in the \heis picture, and quantum Markovianity, as characterized in terms of process tensors \cite{Pollock2018a, Pollock2018}, can be two physically and operationally distinct concepts.

Future work will be devoted to further explore the intricate connection between divisibility in the two pictures.
We aim to derive counterparts of other quantifiers of non-Markovianity in the \schro picture to the \heis picture, both operational and non-operational ones \cite{Budini2022}, connect it to the presence of classical or quantum memory \cite{Giarmatzi2021, Banacki2023, Backer2024}, and causal or noncausal revival of information \cite{Buscemi2025}, as well as exploring whether \heis divisibility is equivalent to monotonic decrease of information \cite{Buscemi2016}.
We also aim to connecting \heis indivisibility to concrete enhancement in tasks hindered by noise.

\section*{Acknowledgments}
  AS and BV acknowledge support from MUR and Next Generation EU via the NQSTI-Spoke1-BaC project QSynKrono (contract n.
  PE00000023-QuSynKrono) and the PRIN 2022 project
Quantum Reservoir Computing (QuReCo) (contract n. 2022FEXLYB).
FS acknowledges support from Magnus Ehrnroothin S\"a\"ati\"o.
DC was supported by the Polish National Science Center under project No. 2018/30/A/ST2/00837.
The authors thank the Toru\'n group and the Aleksander Jab\l o\'nski Foundation for hospitality received.

% ------------------------------------- Biblio -------------------------------------
%\bibliographystyle{quantum}
%\bibliographystyle{unsrtnat}
%\bibliography{biblio}
\bibliography{biblio}

% ------------------------------------- Appendix -------------------------------------
\appendix
% ------------------------------------- P guess Heis -------------------------------------
\section{Proof of Eq.~\eqref{eq:P_guess_H}}
\label{app:proof_P_guess_H}
Consider two effects $0\le E,F\le\id$ and a state $\rho$.
Each effect gives rise to a two-valued probability distribution (the probability of obtaining a `yes' or `no' outcome to the associated measurement) via
\begin{equation}
    p_E \coloneqq \left(p_E(y),\,p_E(n)\right) = \left(\tr[E\,\rho],\,1-\tr[E\,\rho]\right)
\end{equation}
and similarly for $F$.
The probability of committing an error when guessing from which probability distribution the observation is from is given by
\cite{Fuchs1999}
\begin{equation}
    P_{\text{err}}(p_E,p_F\vert\rho) = \frac12\sum_{x\in y,n}\min\{p_E(x),\,p_F(x)\}.
\end{equation}
Suppose, without loss of generality, that $p_F(y)\le p_E(y)$, which implies $p_F(n)\ge p_E(n)$, then
\begin{equation}
    \begin{split}
        P_{\text{err}}(p_E,p_F\vert\rho) &= \frac12\left(p_F(y)+p_E(n)\right) \\&= \frac12-\frac12\tr[(E-F)\rho].
    \end{split}
\end{equation}
The other case $p_F(y)\ge p_E(y)$ just corresponds to swapping $E$ and $F$, and therefore
\begin{equation}
    P_{\text{err}}(p_E,p_F\vert\rho) =  \frac12-\frac12\Big\lvert\tr\big[(E-F)\rho\big]\Big\rvert.
\end{equation}
Using the optimal strategy corresponds to minimizing the error probability, giving 
\begin{equation}
    \begin{split}
        P_{\text{err}}(E,F)&=\inf_\rho P_{\text{err}}(p_E,p_F\vert\rho) \\&= \frac12-\frac12\sup_\rho\Big\lvert\tr\big[(E-F)\rho\big]\Big\rvert \\
        &= \frac12-\frac12\sup_\rho\tr\big[\abs{E-F}\rho\big] \\&= \frac12-\frac12D_\infty(E,F).
    \end{split}
\end{equation}
Thus, the probability of correctly guessing whether $E$ or $F$ was measured is
\begin{equation}
    \begin{split}
        P_{\text{guess}}^{\text{e}}(E,F) &= 1-P_{\text{err}}(E,F)\\&=\frac12\left(1+D_\infty(E,F)\right).
    \end{split}
\end{equation}

% ------------------------------------- Revival bound OD -------------------------------------
\section{Proof of Eq.~\eqref{eq:revival_OD_bound}}
\label{app:revival_OD_bound}

The proof of the information bound for $D_1$ used in \cite{Laine-info-bound,Amato2018a,Campbell2019b} applies to any distance $D(X,Y) = \norm{X-Y}$ provided that the norm $\norm\cdot$ is contracting under the action of completely positive trace preserving transformations, and can be further extended to entropic distinguishability quantifiers \cite{Megier2021,Smirne2022}.
For the operator distance $D_\infty$ of Eq.~\eqref{eq:OD}, the contractivity property does not hold in general and in particular when the considered completely positive trace preserving transformation is the partial trace we rather have \cite{Vacchini2025a,Rastegin2012a}
    \begin{equation}
        \| \tr_E A_{SE} \|_{\infty} \leq   d_E  \| A_{S} \|_{\infty},
    \end{equation}
where $d_E$ is the dimension of the Hilbert space of the environment. The relevant bound on the variation of the distinguishability quantifier $D_\infty$ has to be obtained working in the Heisenberg picture, therefore with the dual maps. We first recall that the dual map to the partial trace for the states is given by the completely positive unital transformation \cite{Vacchini2024}
    \begin{equation}
    \label{eq:obs}
        (\tr_E)^*[X_{S}]=X_{S}\otimes \id_{E}, 
    \end{equation}
so that in correspondence to the two distinct effects $X_{S}^1$ and $X_{S}^2$ it is natural to consider the pair of system-environment observables given by $X_{S}^1\otimes \id_{E}$ and 
$X_{S}^2\otimes \id_{E}$. On the other hand, the dual map to the assignment map    
    \begin{equation}
        \mathcal{A}_{\rho_E}[\rho_{S}]=\rho_{S}\otimes \rho_{E}, 
    \end{equation}
that is used to determine the initial system-environment state given an initial system state, is given by \cite{Vacchini2024}
    \begin{equation}
    \label{eq:redobs}
        \mathcal{A}_{\rho_E}^*[X_{SE}]=\tr_{E}\{ \rho_{E} X_{SE}\}, 
    \end{equation}  
providing the relevant connection between a system-environment observable and the corresponding observable for observations on the system only, obtained averaging out the environmental degrees of freedom.
We now define as usual the unitarily evolved system-environment operators as
\begin{equation}
  X_{SE} (t) = U_{SE} (t)^{\dag} X_S \otimes \id_E U_{SE} (t),
\end{equation}
where $U_{SE} (t)$ is the overall unitary evolution operator and the relevant
observables are chosen according to Eq.~\eqref{eq:obs}. We further introduce
\begin{equation}
  \begin{split}
    X_S (t) & =  \mathcal{A}_{\rho_E}^* [X_{SE} (t)] \\
    & = \tr_E \{ \rho_E U_{SE} (t)^{\dag} X_S \otimes \id_E U_{SE} (t)
    \}
  \end{split}
\end{equation}
and correspondingly
\begin{equation}
  \begin{split}
    X_E (t) & =  \mathcal{A}_{\rho_S}^* [X_{SE} (t)] \\
    & = \tr_S \{ \rho_S U_{SE} (t)^{\dag} X_S \otimes \id_E U_{SE} (t)
    \},
  \end{split}
\end{equation}
according to Eq.~\eqref{eq:redobs}. We now exploit contractivity of the
uniform norm with respect to positive unital transformations, so that we have
\begin{equation}
  \begin{split}
    \| X^1_S (t) - X^2_S (t) \|_{\infty} & = \| \mathcal{A}_{\rho_E}^*
    [X^1_{SE} (t)] - \mathcal{A}_{\rho_E}^* [X^2_{SE} (t)] \|_{\infty}
    \\
    & \leq  \| X^1_{SE} (t) - X^2_{SE} (t) \|_{\infty},
  \end{split}
\end{equation}
and further using invariance under unitary transformations of the uniform norm
\begin{equation}
  \| X^1_S (t) - X^2_S (t) \|_{\infty} \leqslant  \| X^1_{SE} (s) -
  X^2_{SE} (s) \|_{\infty} . 
\end{equation}
We then exploit the identity
\begin{equation}
  \begin{split}
    X^1_{SE} (s) -& X^2_{SE} (s)  = X^1_{SE} (s) - X^1_S (s) \otimes
    X^1_E (s)
    \\
    & + X^1_S (s) \otimes X^1_E (s) - X^1_S (s) \otimes X^2_E (s)
    \\
    & + X^1_S (s) \otimes X^2_E (s) - X^2_S (s) \otimes X^2_E (s)
    \\
    & + X^2_S (s) \otimes X^2_E (s) - X^2_{SE} (s)
  \end{split}
\end{equation}
that combined with the triangle inequality and the following property of the
uniform norm
\begin{equation}
  \| A B \|_{\infty} \leqslant \| A \|_{\infty} \| B \|_{\infty}
\end{equation}
leads to
\begin{equation}
  \begin{split}
    \| X^1_S (t) - &X^2_S (t) \|_{\infty}  \leq \| X^1_{SE} (s) -
    X^1_S
    (s) \otimes X^1_E (s) \|_{\infty} \\
    & + \| X^1_E (s) - X^2_E (s) \|_{\infty} \| X^1_S (s) \|_{\infty}
    \\
    & + \| X^1_S (s) - X^2_S (s) \|_{\infty} \| X^2_E (s) \|_{\infty}
    \\
    & + \| X^2_S (s) \otimes X^2_E (s) - X^2_{SE} (s) \|_{\infty} .
  \end{split}
\end{equation}
For any effect $X_S$ we furthermore have, exploiting contractivity of the
uniform norm with respect to positive unital transformations together with
invariance under unitary transformations
\begin{equation}
  \begin{split}
    \| X _S (s) \|_{\infty} & = \| \mathcal{A}_{\rho_S}^* [X_{SE} (t)]
    \|_{\infty} \\
    & \leq  \| X _{SE} (t) \|_{\infty} \\
    & =  \| X_S \otimes \id_E \|_{\infty} \\
    & =  \| X_S \|_{\infty} \\
    & \leq 1.
  \end{split}
\end{equation}
Combining these inequalities we finally obtain
\begin{widetext}
\begin{equation}
  \begin{split}
    \| X^1_S (t) -& X^2_S (t) \|_{\infty} - \| X^1_S (s) - X^2_S (s)
    \|_{\infty}
     \leq  \| X^1_E (s) - X^2_E (s) \|_{\infty} \\
    & + \| X^1_{SE} (s) - X^1_S (s) \otimes X^1_E (s) \|_{\infty}
     + \| X^2_{SE} (s) - X^2_S (s) \otimes X^2_E (s) \|_{\infty}
  \end{split}
\end{equation}
\end{widetext}
and therefore Eq.~\eqref{eq:revival_OD_bound}.

\section{Connection between \heis and \schro divisibility for the classical 2-state system}
\label{app:classical_divs}
We now show that positivity of $r_{1,2}(t)$ of Eqs.~\eqref{eq:r_class}, \eqref{eq:r_class_2} implies the positivity of $\ell_{1,2}(t)$ of Eqs.~\eqref{eq:l_class}, \eqref{eq:l_class_2}.
Consider first the case in which the denominator is positive, i.e. $a(t) + b(t) -1 \ge 0$.
Positivity of $r_{1,2}(t)$ is equivalent to $\dot a(t)\le0$ and $\dot b(t) \le 0$.
Using the fact that $a(t)-1\le0$, it's easy to show that the numerator of $\ell_1(t)$ is also positive:
\begin{equation}
    -w(t)-\dot b(t) = \dot b(t)\big[a(t)-1\big] - \dot a(t) b(t)\ge0,
\end{equation}
since both terms are positive.
Similarly, using $b(t)-1\le0$, also the numerator of $\ell_2(t)$ is positive:
\begin{equation}
    w(t)-\dot a(t) = \dot a(t)\big[b(t)-1\big] - \dot b(t) a(t)\ge0,
\end{equation}
and therefore $\ell_{1,2}(t)\ge0$.

On the other hand, if $a(t) + b(t) -1 < 0$ \heis divisibility is equivalent to $\dot a(t)\ge0$ and $\dot b(t) \ge 0$, and in a similar manner it is easy to show that the numerators of $\ell_{1,2}$ are now negative
\begin{gather}
    \dot b(t)\big[a(t)-1\big] - \dot a(t) b(t)\le0,\\ \dot a(t)\big[b(t)-1\big] - \dot b(t) a(t)\le0,
\end{gather}
but, since the denominator is also negative, also in this case $\ell_{1,2}(t)\ge0$.

\end{document}